\documentclass[12pt]{article}
\usepackage[margin=1in]{geometry}
\usepackage{setspace}
\usepackage{natbib}
\usepackage{hyperref} 
\usepackage{enumitem}
\usepackage{titlesec}

\usepackage{amsmath, amssymb, amsthm}
\allowdisplaybreaks[4]
\usepackage{CJK}
\usepackage{bm}
\usepackage{caption}
\usepackage{graphicx, subfig}
\usepackage{array}
\usepackage{booktabs}
\usepackage{appendix}
\usepackage{multirow} 

\usepackage{color,xcolor}
\usepackage{algorithm, algorithmic}

\usepackage{authblk}  

\usepackage[normalem]{ulem}

\newtheorem{assumption}{Assumption}
\newtheorem{subassumption}{Assumption}[assumption]
\newtheorem{theorem}{Theorem}

\newtheorem{proposition}{Proposition}

\newtheorem{remark}{Remark}

\def\wh{\widehat}
\def\wt{\widetilde}

\title{\bf Asymptotic Properties of the Distributional Synthetic Controls}

\author[1]{Lu Zhang}
\author[2]{Xiaomeng Zhang}
\author[3,1]{Xinyu Zhang\thanks{*Corresponding author. Email: \texttt{xinyu@amss.ac.cn}}}

\affil[1]{School of Management, University of Science and Technology of China, Hefei, China}
\affil[2]{Econometric Institute, Erasmus University Rotterdam, Rotterdam, Netherlands}
\affil[3]{Academy of Mathematics and Systems Science, Chinese Academy of Sciences, Beijing, China}

\date{}

\begin{document}

\maketitle







\begin{abstract}
As an alternative to synthetic control, the distributional Synthetic Control (DSC) proposed by Gunsilius (2023) provides estimates for quantile treatment effect and thus enabling researchers to comprehensively understand the impact of interventions in causal inference. But the asymptotic properties of DSC have not been built. In this paper,  we first establish the DSC estimator's asymptotic optimality in the essence that the treatment effect estimator given by DSC achieves the lowest possible squared prediction error among all potential estimators from averaging quantiles of control units. We then establish the convergence rate of the DSC weights. A significant aspect of our research is that we find the DSC synthesis forms an optimal weighted average, particularly in situations where it is impractical to perfectly fit the treated unit's quantiles through the weighted average of the control units' quantiles. Simulation results verify our theoretical insights.
\end{abstract}

\vspace{1em}

\section{Introduction}\label{s1}

Causal inference is a pivotal undertaking in social science research, with the synthetic control (SC) method, proposed by \citet{abadie2003economic} and \citet{abadie2010synthetic}, serving as a fundamental tool for assessing the causal effects of policies and interventions in settings with observational data.
However, the original method of synthetic controls predominantly focuses on point estimates of causal effects on aggregate units, neglecting the heterogeneity present in the distributional characteristics.
In many cases, researchers and policy makers want to identify causal effects of policy changes on a treated unit at an aggregate level while having access to data at a finer granularity.
A classic example is the evaluation of the effects of minimum wage policies. In this context, the intervention occurs at the state level, yet researchers have access to individual-level data within a state (e.g., \citeauthor{card1994minimum}, \citeyear{card1994minimum}; \citeauthor{neumark2000}, \citeyear{neumark2000}; \citeauthor{dube2019}, \citeyear{dube2019}). These additional data enable the estimation of heterogeneous treatment effects, shedding light on the varied causal impacts of the policy change across different segments of the population within a state.
Considering the distributional characteristics, in a seminal paper, \citet{gunsilius2023distributional} proposes the distributional synthetic control (DSC) estimator. 
The idea of DSC method is to reconstruct the quantile function associated with the treated unit through a weighted average of quantile functions of the control units, and this weighted average is employed to construct the counterfactual quantile function of the treated unit had it not received treatment.

Compared to the classical SC method, DSC method possesses a notable advantage. It is capable of providing estimates for the effects at different quantiles, enabling researchers to comprehensively understand the impact of interventions.
With this distributional information, researchers can estimate the quantile treatment effect (QTE), which offers certain advantages over the average treatment effect (ATE). While ATE often provides a limited perspective on the impact of a treatment, QTE can reveal more comprehensive insights. It is common for a treatment to leave the mean of the outcome distribution unchanged while affecting its dispersion or altering its shape.
This granularity is particularly valuable when the treatment effects are not uniform across the population.  For instance, a policy intervention might have a larger impact on lower-income individuals compared to higher-income ones.  ATE would only show the overall average effect, potentially overlooking the impacts on the lower-income group. In contrast, QTE would reveal these differences by examining the effects at specific points in the distribution, such as the median or the lower and upper quartiles.
Therefore, both academic research and practical applications place great importance on understanding the treatment's impact on the entire distribution of outcomes.
As stated by \citet{tang2020some}, rather than focusing on the ATE, applied economists and policymakers are increasingly interested in the distributional treatment effect or QTE.
In summary, the QTE function serves as a powerful tool for summarizing the causal effect of a treatment or policy on the marginal distribution of the outcome variable of interest.

Although \citet{gunsilius2023distributional} has demonstrated the notable performance of the DSC method and explored its identification, there remain other important properties that warrant further investigation. Therefore, in this paper, we provide the asymptotic properties of the DSC estimator.
First, we establish the DSC estimator's asymptotic optimality, in the sense that it achieves the lowest possible squared prediction error among all potential treatment effect estimators from averaging quantiles of control units, as the number of draws $M$ goes to infinity.
Second, we show that the DSC weight converges to a limiting weight that minimizes the averaged 2-Wasserstein distance of post-treatment periods. Additionally, we quantify the rate of this convergence.
We find that an enhanced fit before and after the treatment both facilitate the convergence of the DSC weight. Moreover, a larger number of control units is linked to a slower convergence rate. 
We also show that increasing $M$ tightens the bound through the term $M^{-1/4}J$. Finally, we provide a data-driven diagnostic by estimating $\xi_t$ from pre-treatment periods.
Additionally, the asymptotic property of the DSC estimator, established in this paper, does not rely on the model structure. In other words, it does not need to assume the DGP of the potential outcomes, our asymptotic property holds in a model-free setup. Thus, our work includes the factor model used in many studies as a special outcome model.
In the synthetic control literature, \citet{zhang2022asymptotic} and \citet{chen2023synthetic} study the large sample properties of SC estimators.

The rest of the paper is organized as follows. Section 2 introduces the DSC estimator and describes the implementation of the DSC method. Section 3 establishes asymptotic properties of the DSC estimator. Section 4 discusses the assumptions required for asymptotic properties. Section 5 reports the results of Monte Carlo experiments. Section 6 draws some conclusions and briefly points to a natural extension of the DSC method, where mixtures of quantile functions are replaced by mixtures of distribution functions; the details of this extension are presented in Appendix F. Technical proofs are given in the Appendix.

\section{DSC method proposed by Gunsilius (2023)}
In order to facilitate a better understanding of the reader, we will provide a detailed introduction of DSC method below. The methodology and writing style we used in this paper are consistent with \citet{gunsilius2023distributional}, and the setup and notation closely resemble the classical synthetic controls approach (\citeauthor{abadie2003economic}, \citeyear{abadie2003economic}; \citeauthor{abadie2010synthetic}, \citeyear{abadie2010synthetic}; \citeauthor{abadie2021using}, \citeyear{abadie2021using}).

We possess data pertaining to a set of $J + 1$ units, with the first unit ($j = 1$) designated as the treated unit and the subsequent units ($j = 2, \ldots, J + 1$) designated as the potential control units.
The observations span $T$ time periods, and $T_{0}~(T_{0} < T)$ represents the last time period before the treatment in unit $j = 1$. Define $\mathcal{T}_0 = \{ 1, \ldots, T_0 \}$ as the pre-intervention or pre-treatment periods and $\mathcal{T}_1 = \{ T_0+1, \ldots, T_0+T_1 \}$ as the post-intervention or post-treatment periods. All vectors are marked in bold in this paper.


Before delving into the DSC method, we will briefly introduce the classical setting in the literature of SC method.
The classical setting focuses on an aggregated outcome, denoted as $Y_{j t}$, observed for each unit $j = 1, \ldots, J+1$ across the time periods $t = 1, \ldots, T$. Potential outcomes are denoted by $Y_{j t, I}$ when unit $j$ is treated at time $t$ and by $Y_{j t, N}$ when no treatment is applied.
The standard assumption in this setting is that the intervention has no effect on the outcome before the treatment period, ensuring $Y_{j t, N}=Y_{j t, I}$ for all units $j$ and all pre-treatment periods $t \in \mathcal{T}_0$.

The treatment effect, $\alpha_{j t} = Y_{j t, I}-Y_{j t, N}$, for unit $j$ at time $t$ is defined, allowing the observable outcome to be expressed in terms of counterfactual notation as $Y_{j t} = Y_{j t, N}+\alpha_{j t} D_{j t}$, where $D_{j t}=1$ if $j=1$ and $t\in \mathcal{T}_1$, 0 otherwise.
In the classical setting, the interest is to estimate the treatment effect $\alpha_{1t}$ of the treated unit for $t\in \mathcal{T}_1$, that is, $\alpha_{1t} = Y_{1 t, I}-Y_{1 t, N}=Y_{1 t}-Y_{1 t, N}$ for $t\in \mathcal{T}_1$.
Thus, the crucial quantity to estimate is $Y_{1 t, N}$ $(t \in \mathcal{T}_1)$, representing the outcome of the treated unit had it not received the treatment in the post-treatment periods.

The distributional setting in \citet{gunsilius2023distributional} is similar to the classical setting, but with the quantile function $F_{Y_{j t}}^{-1}$ of $Y_{j t}(q)$ as the quantity of interest. The quantile function is formally defined as
$$F^{-1}(q):= \inf _{y \in \mathbb{R}}\{F(y) \geq q\}, \quad q \in(0,1),$$
where $F(y)$ is the corresponding cumulative distribution function.

Analogous to the classical setting, the quantiles of potential outcomes are denoted by $F_{Y_{j t, I}}^{-1}(q)$ when unit $j$ is treated at time $t$ and by $F_{Y_{j t, N}}^{-1}(q)$ when no treatment is applied.
We define $\alpha_{1 t, q} = F_{Y_{1 t, I}}^{-1}(q) - F_{Y_{1 t, N}}^{-1}(q)$ as the treatment effect of the treated unit for each quantile $q~(q\in (0,1))$ in the DSC setting, corresponding to $\alpha_{1 t}$ at the classical SCM setting, and $\widehat{\alpha}_{1 t, q}(\mathbf{w}) = F_{Y_{1 t, I}}^{-1}(q) - \widehat{F}_{Y_{1 t, N}}^{-1}(q)$ is the estimator of $\alpha_{1 t, q}$,
where $\mathbf{w} = (w_2, \ldots, w_{J+1})^{\top}$ is the weight vector belong to the set
$$\mathcal{H} = \left\{\mathbf{w}=\left(w_2, \ldots, w_{J+1}\right)^{\top} \in[0,1]^J \mid \sum_{j=2}^{J+1} w_j = 1\right\}.$$
The purpose of DSC method is to estimate the counterfactual quantile function $F_{Y_{1 t, N}}^{-1}(q)$ of the treated unit had it not received treatment.
The DSC method constructs the estimation of $F_{Y_{1 t, N}}^{-1}(q)$ through an optimally weighted average of the control quantile functions $F_{Y_{j t}}^{-1}(q), ~j=2, \ldots, J+1$, that is,
$$ \widehat{F}_{Y_{1 t, N}}^{-1}(q) = \sum_{j=2}^{J+1} w_j F_{Y_{j t}}^{-1}(q) \quad \text { for all } q \in(0,1). $$

The question moves to how DSC method determines $\mathbf{w}$. This question will be addressed in the subsequent exposition.
The estimator $\widehat{\mathbf{w}}$ of $\mathbf{w}$ is a weighted average of the weights $\widehat{\mathbf{w}}_{t}$ over all pre-treatment periods, where $\widehat{\mathbf{w}} = \left(\widehat{w}_{2}, \ldots, \widehat{w}_{(J+1)} \right)^{T}$.
Thus, to derive the weight $\widehat{\mathbf{w}}$, we need to first obtain weights $\widehat{\mathbf{w}}_t$ at each time period $t$ ($t \in \mathcal{T}_{0}$), where $\widehat{\mathbf{w}}_t = \left(\widehat{w}_{2t}, \ldots, \widehat{w}_{(J+1)t} \right)^{T}$.
For each time period $t \in \mathcal{T}_{0}$, the weights $\widehat{\mathbf{w}}_t \in \mathcal{H}$ are determined to ensure that the weighted average of quantile functions for the control units closely approximates that of the treated unit.
To quantify the accuracy of approximation mathematically, \citet{gunsilius2023distributional} choose the 2-Wasserstein distance as it can simplify the task of determining the weights $\widehat{\mathbf{w}}_t$ into a straightforward regression problem. Following \citet{gunsilius2023distributional}, the 2-Wasserstein distance, denoted $W_{2}(P_{1}, P_{2})$, between two probability measures $P_{1}$ and $P_{2}$ with finite second moments is defined as
$$ W_2\left(P_1, P_2\right)=\left(\int_0^1\left|F_1^{-1}(q)-F_2^{-1}(q)\right|^2 d q\right)^{1 / 2}, $$
where $F_1^{-1}(q)$ and $F_2^{-1}(q)$ are the quantile functions corresponding to $P_1$ and $P_2$, respectively.

Consequently, the method involves determining the weights to minimize the distance between $\sum_{j=2}^{J+1} \widehat{w}_{jt} F_{Y_{j t}}^{-1}(q)$ and the target $F_{Y_{1 t}}^{-1}(q)$ in the 2-Wasserstein space.
For each $t \in \mathcal{T}_0$, \citet{gunsilius2023distributional} determine the weights by solving
\begin{eqnarray}
\widehat{\mathbf{w}}_{t}^{(1)} = \underset{\mathbf{w}_{t} \in \mathcal{H}}{\operatorname{argmin}} \int_0^1 \left|\sum_{j=2}^{J+1} w_{jt} F_{Y_{j t}}^{-1}(q) - F_{Y_{1 t}}^{-1}(q) \right|^2 dq. \label{3.1}
\end{eqnarray}

The optimization (\ref{3.1}) is a convex problem for the weights $\mathbf{w}_{t}$, guaranteeing a unique solution. In practice, the integral can be approximated by randomly sampling a considerable number $M$ of draws $\left\{V_m\right\}_{m=1}^M$ from the uniform distribution on the unit interval, i.e., $V_m \sim$ $U[0,1]$ and solving
\begin{eqnarray}
\widehat{\mathbf{w}}_{t}^{(2)} = \underset{\mathbf{w}_{t} \in \mathcal{H}}{\operatorname{argmin}} \frac{1}{M} \sum_{m=1}^M \left| \sum_{j=2}^{J+1} w_{j t} F_{Y_{j t}}^{-1}\left(V_m\right) - F_{Y_{1 t}}^{-1}\left(V_m\right) \right|^2. \label{3.02}
\end{eqnarray}
It is worth noting that we allow sequence $\{V_m,~m = 1, \ldots, M\}$ to be dependent, aligning more closely with actual data characteristics, while \citet{gunsilius2023distributional} necessitates their independence.

\begin{remark}[Dependent draws $\{V_m\}_{m=1}^M$ and practical implementations]
    As noted in \cite{gunsilius2023distributional}, drawing $\{V_m\}$ i.i.d.\ from $U[0,1]$ is already sufficient to approximate the integral in the Wasserstein distance and is natural and theoretically well-justified. The relaxation to dependent sequences is not because the i.i.d.\ assumption is inadequate, but brings two practical considerations. 
First, in any real implementation, the draws $\{V_m\}$ are produced by a pseudo-random number generator such as the Mersenne Twister, which is a deterministic recursive map; hence the draws $\{V_m\}$ are not independent in a strict mathematical sense. The dependence of $\{V_m\}$ across $m$ decays rapidly, which is precisely the structure captured by a mixing condition.	
Second, it enables more computational efficiency via quasi-Monte Carlo~(QMC) or Gaussian Quadrature.
QMC sequences such as the Sobol or Halton sequence are determenistic and hence dependent by construction. 
By the Koksma-Hlawka inequality \citep{niederreiter1992random}, the approximation error of QMC satisfies
	\begin{align}
	&\left| \int_{0}^{1} \left| \sum_{j=2}^{J+1} w_{jt} F_{Y_{jt}}^{-1}(q) - F_{Y_{1t}}^{-1}(q) \right|^2  dq  - \frac{1}{M} \sum_{m=1}^{M}\left| \sum_{j=2}^{J+1} w_{jt} F_{Y_{jt}}^{-1}(V_m) - F_{Y_{1t}}^{-1}(V_m) \right|^2
	\right|
	\nonumber\\
	\leq& V\left(\left| \sum_{j=2}^{J+1} w_{jt} F_{Y_{jt}}^{-1}(q) - F_{Y_{1t}}^{-1}(q) \right|^2\right) D^*_M,
	\nonumber
	\end{align}
where $V(f)$ is the total variation of the real-valued function $f$ and $D^*_M$ is the star discrepancy of the sequence $\{V_m\}_{m=1}^{M}$. 
We know that $V\left(\left| \sum_{j=2}^{J+1} w_{jt} F_{Y_{jt}}^{-1}(q) - F_{Y_{1t}}^{-1}(q) \right|^2\right)$ is generally bounded  in many empirical applications, especially when the outcome variable has bounded support, such as wages, test scores, or market shares; in addition, $D^*_M = O(\log M / M)$ for low-discrepancy sequences when the dimension of $V_m$ is one, compared to $O(M^{-1/2})$ for i.i.d.\ draws.
When the quantile function $F_{Y_{jt}}^{-1}(q)$ is sufficiently smooth, applying Gaussian Quadrature (based on Gaussian nodes) to approximate the integral in the Wasserstein distance can achieve significantly higher computational efficiency than using i.i.d. draws from $U[0,1]$.
Therefore, replacing i.i.d. draws with either a QMC sequence or Gaussian Quadrature obtains more accurate estimates of DSC weights  for the same value of $M$ at no additional computational cost. 
We note that fully deterministic QMC sequences and Gaussian nodes sit at the boundary of what a mixing condition formally covers; we mention them here to illustrate the spirit of the generalization. 
\end{remark}

When the quantile functions $F_{Y_{j t}}^{-1}$ are known, it becomes possible to construct an artificial sample $\widetilde{Y}_{j t m} = F_{Y_{j t}}^{-1}(V_m)$ indexed by $m$ with the number of draws $M$, where the choice of $M$ in the approximation is determined by the researcher. Analogously, the quantiles of potential outcomes are denoted by $\widetilde{Y}_{j t m, I}$ when unit $j$ is treated at time $t$ and by $\widetilde{Y}_{j t m, N}$ when no treatment is applied.
In practice, however, the quantile functions $F_{Y_{j t}}^{-1}(q)$ are unknown and necessitate estimated from available data. The empirical quantile functions $\widehat{F}_{Y_{j t n_{j}}}^{-1}(q)$, based on the samples
$\{Y_{l,jt}\}_{l=1}^{n_j}$ for $j=1,\ldots,J+1$ and $t\in\{1,\ldots,T\}$, are used as the estimator of $F_{Y_{j t}}^{-1}(q)$, where the subscript $n_{j}$ in $\widehat{F}_{Y_{j t n_{j}}}^{-1}(q)$ denotes that it is based on $n_{j}$ samples. We view $\{Y_{l,jt}\}_{l=1}^{n_j}$ as a sample from the unit--time outcome distribution $F_{Y_{jt}}$.
An approach commonly employed for this estimation is through order statistics: $\widehat{F}_{Y_{j t n_{j}}}^{-1}(q) = Y_{t, n_{j} (k)}$, where $k$ is selected such that $(k-1)/n_{j} < q < k/n_{j}$, $Y_{t, n_{j} (k)}$ represents the order statistics of the sample $\{Y_{l, j t}\}, l = 1, \ldots, n_{j}$, $j = 1, \ldots, J+1$, and subscript $n_{j}(k)$ represents the $k$-th sample of the samples $\{Y_{l,j t}\}_{l=1}^{n_j}$ after sorting $\{Y_{l, j t}\}_{l=1}^{n_j}$ in ascending.
Correspondingly, in practice, we let $\widehat{\widetilde{Y}}_{j t m} = \widehat{F}_{Y_{j t n_{j}}}^{-1}\left(V_m\right)$ and choose $M$ such that $M=Cn$, where $C$ is a constant and $n=\min\{n_{1}, n_{2}, \ldots, n_{J+1}\}$.

We define the loss function for each $t \in \mathcal{T}_0$ as $L_{t}(\mathbf{w}_t) = M^{-1} \sum_{m=1}^M |\sum_{j=2}^{J+1} w_{j t} \widehat{\widetilde{Y}}_{j t m} - \widehat{\widetilde{Y}}_{1 t m} |^2$.

One can then write the expression (\ref{3.02}) as a linear regression, that is,
\begin{eqnarray}
\widehat{\mathbf{w}}_{t} &=& \underset{\mathbf{w}_{t} \in \mathcal{H}}{\operatorname{argmin}} L_{t}(\mathbf{w}_t) \nonumber\\
&=& \underset{\mathbf{w}_{t} \in \mathcal{H}}{\operatorname{argmin}} \frac{1}{M} \sum_{m=1}^M\left|\sum_{j=2}^{J+1} w_{j t} \widehat{\widetilde{Y}}_{j t m} - \widehat{\widetilde{Y}}_{1 t m}\right|^2 \nonumber\\
&=& \underset{\mathbf{w}_{t} \in \mathcal{H}}{\operatorname{argmin}}\left\|\widehat{\widetilde{\mathbb{Y}}}_t \mathbf{w}_t - \widehat{\widetilde{\boldsymbol{Y}}}_{1 t} \right\|_2^2, \label{3.2}
\end{eqnarray}
where $\widehat{\widetilde{\mathbb{Y}}}_t$ is the $M \times J$-matrix with entry $\widehat{\widetilde{Y}}_{(j+1) t m}$ at position $(m, j)$, $\widehat{\widetilde{\boldsymbol{Y}}}_{1 t}$ is the vector of elements $\widehat{\widetilde{Y}}_{1 t m}$ for $m=1, \ldots, M$, and $\|\cdot\|_2$ denotes the Euclidean norm on $\mathbb{R}^M$.

Subsequently, the DSC weight $\widehat{\mathbf{w}}$ can be calculated as a weighted average of the weights $\widehat{\mathbf{w}}_{t}$ over all pre-treatment periods, that is,

$$\widehat{\mathbf{w}} = \sum_{t \in \mathcal{T}_0} \lambda_{t} \widehat{\mathbf{w}}_{t} \quad \text{for } \lambda_{t}\geq 0 \text{ and } \sum_{t \in \mathcal{T}_0} \lambda_{t} = 1.$$
Regarding the choice of weights $\lambda_{t}$, viable options are provided by \citet{arkhangelsky2021synthetic}, which are also applicable in this case.
At every time point $t \in \mathcal{T}_{1}$ within the post-treatment period, the counterfactual quantile function for the treated unit had it not received the treatment is calculated by $\widehat{F}_{Y_{1t n_1, N}}^{-1} = \sum_{j=2}^{J+1} \widehat{w}_{j} \widehat{F}_{Y_{j t n_{j}}}^{-1}$.


In summary, the algorithm for DSC method is shown in Algorithm 1.

\begin{algorithm}
	\renewcommand{\algorithmicrequire}{\textbf{Input:}}
	\caption{Distributional Synthetic Controls.}
	\label{alg:1}
	\begin{algorithmic}[1]
		\REQUIRE 1. data $Y_{l, j t}$ with $l=1, \ldots, n_j, j=1, \ldots, J+1, t=1, \ldots, T$\\
                 ~~~~~~2. weights $\left\{\lambda_{t}\right\}_{t \in \mathcal{T}_0}$ that satisfy $\lambda_{t}\geq 0$ and $\sum_{t \in \mathcal{T}_0} \lambda_{t} = 1$
		\STATE \textbf{procedure} DSC
		\FOR{each time period $t \in \mathcal{T}_{0} \cup \mathcal{T}_{1}$}
        \FOR{each unit $j=1, \ldots, J+1$}
		\STATE estimate the empirical quantile functions $\widehat{F}_{Y_{j t n_j}}^{-1}$
        \ENDFOR
		\ENDFOR
        \FOR{each time period $t \in \mathcal{T}_0$}
		\STATE obtain the weights $\widehat{\mathbf{w}}_{t}$ by solving (\ref{3.1}) via the regression 
        $$\widehat{\mathbf{w}}_{t} = \underset{\mathbf{w}_{t} \in \mathcal{H}}{\operatorname{argmin}} \|\widehat{\widetilde{\mathbb{Y}}}_t \mathbf{w}_t - \widehat{\widetilde{\boldsymbol{Y}}}_{1 t} \|_2^2$$
        \ENDFOR
        \STATE obtain the DSC weight $\widehat{\mathbf{w}} = \sum_{t \in \mathcal{T}_0} \lambda_{t} \widehat{\mathbf{w}}_{t}$ over all $t \in \mathcal{T}_0$
        \FOR{each time period $t\in \mathcal{T}_1$}
		\STATE obtain the estimation of the counterfactual quantile function $\widehat{F}_{Y_{1 t n_1, N}}^{-1} = \sum_{j=2}^{J+1} \widehat{w}_j \widehat{F}_{Y_{j t n_j}}^{-1}$
        \ENDFOR
		\STATE \textbf{end procedure}
	\end{algorithmic}
\end{algorithm}

\section{Assumptions and main results}
In this section, we will list some assumptions and present our theoretical results. Our first result is the asymptotic optimality of the DSC estimator in the sense that it achieves the lowest possible averaged 2-Wasserstein distance of post-treatment periods among all possible averaging estimators over control units, when the $M$ goes to infinity. Additionally, we establish the convergence of DSC weights to the infeasible optimal weights that minimize the averaged 2-Wasserstein distance of post-treatment periods. Unless specified otherwise, all limiting properties hold as $M \rightarrow \infty$.

To facilitate the theoretical analysis, we define the corresponding risk function for each $t \in \mathcal{T}_0$ as
$R_{t}(\mathbf{w}_t) = \mathbb{E}_{V_m} [L_{t}(\mathbf{w}_t) \mid \mathcal D]$, 
where $\mathcal D \equiv \{Y_{l,jt}: l=1,\ldots,n_j;\ j=1,\ldots,J+1;\ t=1,\ldots,T\}$ denotes the collection of observed samples. 
Note that randomness in our setup arises from two sources: the draws $\{V_m\}_{m=1}^M$ used to approximate the integral, and the observed sample $\mathcal D$ used to construct the empirical quantile functions. 
To avoid ambiguity, we formulate risks conditionally on $\mathcal D$ and take expectations only with respect to $V_m$.
For notational convenience, throughout the paper, we use the shorthand $\mathbb E[\cdot]\equiv \mathbb E_{V_m}[\cdot\mid \mathcal D]$, i.e., $\mathbb E[\cdot]$ denotes expectation with respect to the $V_m$ conditional on $\mathcal D$. Then $R_t(\mathbf w_t)$ can be written equivalently as
\[
R_{t}(\mathbf{w}_t) = \frac{1}{M} \sum_{m=1}^M \mathbb{E} \left|\sum_{j=2}^{J+1} w_{j t} \widehat{\widetilde{Y}}_{jtm}
- \widehat{\widetilde{Y}}_{1tm}\right|^2.
\]

To evaluate the performance of the DSC estimator, we consider the average of the 2-Wasserstein distance at post-treatment period for some weight $\mathbf{w} \in \mathcal{H}$, defined as
$$\bar{R}_{T_1}(\mathbf{w}) = \frac{1}{T_1} \sum_{t \in \mathcal{T}_1} \int_0^1 \left( \sum_{j=2}^{J+1} w_{j} F_{Y_{j t}}^{-1}(q) - F_{Y_{1 t, N}}^{-1}(q) \right)^2 dq.$$

Define $\xi_{t} = \inf _{\mathbf{w_t} \in \mathcal{H}} R_{t}(\mathbf{w}_{t})$, and $\bar{\xi}_{T_1}=\inf _{\mathbf{w} \in \mathcal{H}} \bar{R}_{T_1}(\mathbf{w})$.
To show the asymptotic optimality and convergence, the following assumptions are used. All explanations of these assumptions are given in the next section.

\begin{assumption}
$\xi_{t}^{-1} M^{-1/2} J^{2} = o(1)$ for $t \in \mathcal{T}_0$.
\end{assumption}

\begin{assumption}
$\sup _{\mathbf{w}_{t} \in \mathcal{H}} \left| R_{t}(\mathbf{w}_t) - \bar{R}_{T_1}(\mathbf{w}_t) \right| = O(n^{-1/2} J^2 ) + o(\xi_{t})$ for $t \in \mathcal{T}_0$.
\end{assumption}

\begin{assumption}
For each fixed $i \in\{1, \ldots, J+1\}$ and $t \in \mathcal{T}_0 \cup \mathcal{T}_1$, $\{\widehat{\widetilde{Y}}_{itm, N}\}$, the sequence indexed by $m$, is either $\alpha$-mixing with the mixing coefficient $\alpha = -r/(r-2)$ or $\phi$-mixing with the mixing coefficient $\phi = -r /(2 r-1)$ for $r \geq 2$.
\end{assumption}

Denote $e_{t, m, \widehat{\widetilde{Y}}_N}^{(i)} = \widehat{\widetilde{Y}}_{itm,N} - \widehat{\widetilde{Y}}_{1tm,N}$ for $i \in\{2, \ldots, J+1\}$, $m \in \{1, 2, \ldots, M\}$ and $t \in \mathcal{T}_0 \cup \mathcal{T}_1$.

\begin{assumption}

(i) There exists a constant $C_1$ such that $\mathbb{E} \{\widehat{\widetilde{Y}}_{itm, N}\}^4 \leq C_1 < \infty$ for $i \in\{1, \ldots, J+1\}$, $m \in \{1, 2, \ldots, M\}$ and $t \in \mathcal{T}_0 \cup \mathcal{T}_1$.

(ii) There exists a constant $C_2$ such that $\operatorname{var} ( M^{-1 / 2} \sum_{m=1}^M e_{t, m, \widehat{\widetilde{Y}}_N}^{(i)} e_{t, m, \widehat{\widetilde{Y}}_N}^{(j)} ) \geq C_2 > 0$ for $M$ sufficiently large, and for any $i, j \in\{2, \ldots, J+1\}$ and $t \in \mathcal{T}_0 \cup \mathcal{T}_1$.
\end{assumption}

\begin{theorem}\label{th1}
\it Given any $\mathbf{\lambda}$, if $T_0$ is finite, then under Assumptions 1-4, we have
\begin{align}
\frac{\bar{R}_{T_1}(\widehat{\mathbf{w}})}{\inf _{\mathbf{w} \in \mathcal{H}} \bar{R}_{T_1}(\mathbf{w})} \stackrel{p}{\rightarrow} 1 . \label{4.2}
\end{align}
\end{theorem}

Theorem \ref{th1} establishes the asymptotic optimality of the DSC estimator.
Specifically, (\ref{4.2}) shows that the DSC weight is asymptotically optimal among all possible weighting combinations in the sense that the averaged 2-Wasserstein distance of the DSC estimator is asymptotically identical to those of the infeasible but best estimator. Moreover, the result in Theorem \ref{th1} can be understood as consisting of two conceptually distinct steps. The first step is a purely statistical result establishing the asymptotic optimality of the proposed estimator with respect to the pre-treatment risk. The second step relies on the identification condition in Assumption 2, which links the pre-treatment and post-treatment risks and allows the post-treatment optimality result in Theorem \ref{th1} to be derived from the pre-treatment optimality result. A more detailed discussion of this decomposition is provided in Appendix A.2.

This optimality statement is closely related to the classical SCM theory of \citet{zhang2022asymptotic}, who establish an analogous risk-optimality property under the mean-squared prediction error (MSPE) criterion for aggregate-level outcomes. Our result can be viewed as a distributional analogue: we evaluate prediction performance through the squared $2$-Wasserstein distance between outcome distributions, and show that the DSC weights achieve asymptotic optimality. Moreover, in the distributional setting, the loss depends on estimated quantile functions and hence incorporates within-unit sampling variability. Accordingly, our limiting argument is driven by $M\to\infty$ (allowing $T_0$ to be fixed), rather than relying primarily on a long pre-treatment time series. This regime is natural for applications where each unit-time cell contains many micro-level observations, and it clarifies that DSC remains asymptotically optimal even when the number of pre-treatment periods is limited.

We define $\mathbf{\Sigma}_t = M^{-1} \mathbb{E}(\widehat{\widetilde{\mathbb{Y}}}_t^{\top} \widehat{\widetilde{\mathbb{Y}}}_t)$ for $t \in \mathcal{T}_0$, where $\mathbb E[\cdot]$ here denotes expectation over $V_m$ conditional on $\mathcal D$, and use $\lambda_{min}(\cdot)$ and $\lambda_{max}(\cdot)$ to represent the minimum and maximum eigenvalue of a matrix.


\begin{assumption}
There exist constants $\kappa_{1}$ and $\kappa_{2}$ such that $0 < \kappa_{1} \leq \lambda_{min}(\mathbf{\Sigma}_t) \leq \lambda_{max}(\mathbf{\Sigma}_t) \leq \kappa_{2}$ for $t \in \mathcal{T}_0$.
\end{assumption}



The optimal weight vector for a given $T_1$ is defined as the minimizer of the $\bar{R}_{T_1}(\mathbf{w})$, i.e.,
\begin{align}
\mathbf{w}_{T_1}^{\text {opt}} = \underset{\mathbf{w} \in \mathcal{H}}{\arg \min} \bar{R}_{T_1}(\mathbf{w}).
\label{4.002}
\end{align}

\begin{theorem}\label{th2}

Given any $T_1$, if $\mathbf{w}_{T_1}^{\text {opt }}$ is an interior point of $\mathcal{H}$ and Assumptions 2 and 5 hold, then
\begin{align}
\left\|\widehat{\mathbf{w}}-\mathbf{w}_{T_1}^{opt}\right\| = O_p \left(\bar{\xi}^{1/2} + \bar{\xi}_{T_1}^{1/2} + M^{-1/4} J\right), \label{4.003}
\end{align}
where $\bar{\xi} = \sum_{t \in \mathcal{T}_0} \lambda_{t} \xi_{t}$,
with any given weight $\lambda_{t}~(t\in \mathcal{T}_0)$ satisfying $\lambda_{t}\geq 0$ and $\sum_{t \in \mathcal{T}_0} \lambda_{t} = 1$.
\end{theorem}

Theorem 2 provides a bound for the Euclidean norm of the difference between the DSC weight and the infeasible optimal weight. If we aim to make the bound on the right-hand side of equation (\ref{4.003}) converge to 0, it is not only necessary for $M \rightarrow \infty$, but also for both $\xi_{t}$ for $t \in \mathcal{T}_0$ and $\bar{\xi}_{T_1}$ to tend towards 0; therefore, if these requirements are satisfied, Theorem 2 will establish both the convergence of DSC weight $\widehat{\mathbf{w}}$ to the infeasible optimal weight $\mathbf{w}_{T_1}^{\text {opt}}$ and quantifies the rate of convergence, which depends on $\bar{\xi}$, $\bar{\xi}_{T_1}$ and $J$ as $M\rightarrow \infty$. We discuss the roles of $\bar{\xi}$, $\bar{\xi}_{T_1}$, $J$ and $M$ in turn.
First, a faster rate of $\bar{\xi}$ and $\bar{\xi}_{T_1}$ going to zero implies quicker convergence of $\widehat{\mathbf{w}}$. Since $\bar{\xi}$ is the weighted average of $\xi_{t}~(t \in \mathcal{T}_0)$, where $\xi_{t}$ serves as a measure of the fit of the quantiles of the control units to the quantile of the treated unit for each $t \in \mathcal{T}_0$, the theorem establishes a link between good pre-treatment fit and accurate weight estimation.
Then, from the term $M^{-1/4} J$, we know that a larger $J$ is linked to a slower convergence rate.
Anticipated reductions in estimation accuracy are expected with an increase in the dimension of parameters, given that $J$ corresponds to the number of weight parameters to be estimated.
Finally, the role of $M$ is transparent from the term $M^{-1/4}J$: holding $J$ fixed, increasing $M$ decreases this component at rate $M^{-1/4}$.

Theorem 2 is also connected to the weight-convergence theory for classical SCM developed in \citet{zhang2022asymptotic}. Under an MSPE criterion for aggregate-level outcomes, they provide an explicit upper bound for $\|\widehat{\mathbf w}-\mathbf w^{\mathrm{opt}}_{T_1}\|$ in terms of pre- and post-treatment approximation errors and a term that increases with the donor dimension $J$, with the asymptotic argument primarily driven by the length of the pre-treatment period $T_0$. Our Theorem 2 provides a distributional analogue under the squared $2$-Wasserstein loss: the approximation errors $\bar\xi$ and $\bar\xi_{T_1}$ are defined via quantile-function fitting, and the relevant sample-size dimension is governed by $M$, the within-unit sample size underlying empirical quantiles. Accordingly, Theorem 2 shows that consistent estimation of the DSC weights requires not only a sufficiently large within-unit sample size, but also vanishing approximation errors, so that the treated-unit distribution can be well approximated by convex combinations of donor distributions both in the pre-treatment period and for the post-treatment counterfactual.

\section{Discussion of the assumptions}
This section discusses Assumptions 1-5 and their roles in our theoretical analysis.  
Since Assumption 2 is stated at a relatively high level, we provide model-based illustrations under concrete and stylized settings to aid interpretation.
These conditions are presented as interpretable sufficient conditions and are not additional assumptions required for our main results.

Assumption 1 imposes restrictions on the relative rate of several quantities approaching infinity, i.e., $\xi_{t},~M$ and $J$.
It is crucial to highlight that this assumption implies $\xi_{t} \neq 0$, a crucial assumption for establishing the asymptotic optimality of the DSC weight. Intuitively, $\xi_{t} \neq 0$ means that for each $t \in \mathcal{T}_0$, it is impossible to achieve a perfect fit for the pre-treatment quantiles of the treated unit using a linear combination of the quantiles of the control units, and this situation is referred to as an imperfect pre-treatment fit.
Assumption 1 can be connected to the classical SC setting (e.g. \citeauthor{abadie2010synthetic}, \citeyear{abadie2010synthetic}), as it mirrors the requirement in the classical SC setting to obtain a unique set of weights. Specifically, if $\xi_t = 0$, then the pre-treatment quantiles of the treated unit lie within the (geodesic) convex hull of the quantiles of the control units. By Carathéodory’s theorem, $\xi_t = 0$ implies that there may exist multiple sets of weights that achieve an exact fit, resulting in the DSC weights failing to be identified. By ruling out this perfect fit scenario, \(\xi_t>0\) ensures the uniqueness of the optimal DSC weight.

Furthermore, $\xi_t$ plays a pivotal role in determining the practical applicability of the theoretical properties, as Assumption 1 -- required for guarantees -- assumes that $\xi_t>0$. The magnitude of $\xi_t$ directly influences the convergence rate of the weight $\widehat{\mathbf{w}}$, making it relevant for both theoretical and empirical considerations. 
Importantly, in practice, \(\xi_t\) can be estimated from pre-treatment data. This provides a simple, data-driven diagnostic for assessing whether the condition implicit in Assumption 1 is plausible in a given application. We recommend approximating it using the sample analogue
\begin{align}
    \widehat\xi_t = \min_{\mathbf{w_t} \in \mathcal{H}} \frac{1}{M} \sum_{m=1}^M (\sum_{j=2}^{J+1} w_{j t} \widehat{\widetilde{Y}}_{jtm} - \widehat{\widetilde{Y}}_{1tm})^2. \label{xi-hat}
\end{align}
This diagnostic enables researchers to empirically learn about the magnitude and behavior of $\xi_t$. 
Therefore, reporting $\widehat\xi_t$ can be informative in empirical work as a complementary diagnostic alongside standard pre-treatment fit measures.

We next relate Assumption 1 to the implementation choice of $M$ used in practice, which clarifies that the assumption effectively imposes a requirement on within-unit sample sizes.
Assumption 1 is stated in terms of $(\xi_t,M,J)$, whereas our empirical implementation fixes $M$ to be proportional to the minimal within-unit sample size. Specifically, we choose $M=Cn$. Under this choice, Assumption 1 is equivalently rewritten as the sample-size condition
$$\xi_t^{-1} n^{-1/2} J^{2} = o(1)~ for~ t \in \mathcal{T}_0.$$
Hence, in the pre-treatment imperfect-fit regime $\xi_t>0$, Assumption 1 imposes a restriction on the sample size. 
This is natural because, while $M$ controls the discretization used to approximate the integral representation of the Wasserstein loss, the quantile functions $F^{-1}_{Y_{jt}}(q)$ are unknown and must be estimated from $n_j$ individual-level observations. Accordingly, $\widehat F^{-1}_{Y_{jtn_j}}(q)$ is an empirical quantile function whose effective information content is governed by $n_j$. Consequently, choosing $M$ far larger than $n$ mainly re-samples essentially the same set of order statistics more densely and increases computational cost, without delivering commensurate additional information. 
These considerations motivate letting $M$ grow proportionally with $n$, i.e., $M=Cn$, which aligns the discretization level with the effective sample size while keeping computation manageable.

Assumption 2 constrains the difference between the fits in each pre-treatment period $t$ and the fits in the post-treatment periods. This implies that the primary distinction between the quantiles of the outcomes at each pre-treatment period $t$ and post-treatment period is entirely attributable to the treatment effect.
A similar assumption has been discussed in \citet{hansen2012jackknife}.

In the distributional setting, the quantity we are interested in is the quantile function $F_{Y_{j t}}^{-1}$ of $Y_{j t}$ instead of $Y_{j t}$. To provide sufficient conditions under which Assumption 2 holds, we consider a stylized quantile factor structure similar to that studied in \citet{chen2021quantile} to generate the potential outcome of the quantile version. This quantile factor representation serves as a quantile-level analogue of the traditional factor models commonly employed in the SCM literature, and is introduced purely as an analytical device rather than as a structural model of distributional dynamics.
Specifically, suppose that the potential outcomes $\widetilde{Y}_{itm, N}$ are generated from the following quantile factor model:
\begin{align}
\widetilde{Y}_{itm, N} = Q_{Y_{it, N}}[v_{m} | \boldsymbol{f}_{t, m}] = \boldsymbol{\lambda}_{i, m}^{\top} \boldsymbol{f}_{t, m},\label{m.1}
\end{align}
where the subscript $m$ has the same meaning as in Section 2, and $v_{m}$ is the observation of $V_{m}$, $\boldsymbol{f}_{t, m} $ signifies an $F_{m} \times 1$ vector of unobserved random common factors, $\boldsymbol{\lambda}_{i,m}$ is an $F_{m} \times 1$ vector of non-random factor loadings.

Since the quantile function $F_{Y_{j t}}^{-1}$ is estimated using the order statistic, we have $\widehat{F}_{Y_{j t n_{j}}}^{-1}(V_{m}) = Y_{t, n_{j} (k)}$.
Let $V_{m}^{*} \in (0,1)$ denote the population quantile level such that $F_{Y_{j t}}^{-1}\left(V_m^*\right) = Y_{t, n_{j} (k)}$. That is, $V_{m}^{*}$ is the true quantile level at which the sample quantile $Y_{t, n_{j} (k)}$ lies.
Note that $V_{m}^{*}$ depends on $m$, and converges to $V_m$ as the sample size increases. To simplify notation, 
we write $F_{Y_{j t}}^{-1}\left(V_m^*\right) = \widetilde{Y}_{j t m^{*}}.$
Accordingly, $\widehat{\widetilde{Y}}_{itm, N}$ can be expressed in the form of a factor model:
$$\widehat{\widetilde{Y}}_{itm, N} = \boldsymbol{\lambda}_{i, m^{*}}^{\top} \boldsymbol{f}_{t, m^{*}}.$$


Then, Assumption 2 can be derived from more general assumptions as follows. The detailed proof is provided in Appendix C.

\setcounter{assumption}{2}
\begin{subassumption}
We treat $\{\boldsymbol{\lambda}_{i, m}~|~i\in \{1,2, \ldots, J+1 \}, ~m \in \{1, 2, \ldots, M\} \}$ as fixed and $\{\boldsymbol{f}_{t,m}~|~m\in \{1,2,\ldots, \ldots M \}, ~t \in \mathcal{T}_0 \cup \mathcal{T}_1 \}$ as stochastic.
\end{subassumption}

\begin{subassumption}
\noindent $T_1^{-1} \sum_{t_1 \in \mathcal{T}_1} \mathbb{E}\left( M^{-1} \sum_{m=1}^M \boldsymbol{f}_{t,m^{*}}^{\top} \boldsymbol{f}_{t,m^{*}} - M^{-1} \sum_{m=1}^M \boldsymbol{f}_{t_1, m^{*}}^{\top} \boldsymbol{f}_{t_1, m^{*}} \right) = O (n^{-1/2})$ for $t \in \mathcal{T}_0$.
\end{subassumption}

\begin{subassumption}
$\boldsymbol{\lambda}_{i, m}$ and $\boldsymbol{\lambda}_{i, m^{*}}$ are bounded uniformly for $i\in \{1, 2, \ldots, J+1\}$ and $m\in\{1, 2, \ldots, M\}$.
\end{subassumption}

\begin{subassumption}
\noindent $T_1^{-1} \sum_{t \in \mathcal{T}_1} \mathbb{E}\left( M^{-1} \sum_{m=1}^M \boldsymbol{f}_{t,m}^{\top} \boldsymbol{f}_{t,m} - M^{-1} \sum_{m=1}^M \boldsymbol{f}_{t, m^{*}}^{\top} \boldsymbol{f}_{t, m^{*}} \right) = O ( n^{-1/2})$.
\end{subassumption}

Assumption 2.1 serves to simplify the proof, and a similar assumption is also employed in \citet{ferman2021synthetic} and \citet{ferman2021properties}.
Assumption 2.2 implies that for each $t \in \mathcal{T}_0$, the common factors may differ from the average of the common factors across all post-treatment periods. However, this discrepancy gradually diminishes as more samples are used, indicating that the variation of the common factors should not vary significantly after treatment. This assumption means that the treatment effect alone accounts for the majority of the difference between the
quantiles of the outcomes at each pre-treatment period $t$ and post-treatment period.
Assumption 2.3 requires the uniform boundedness of the factor loadings, and the same assumption is also employed in \citet{ferman2021properties}.
Assumption 2.4 requires that there be no substantial discrepancy between the common factors $\boldsymbol{f}_{t,m}$ and $\boldsymbol{f}_{t, m^{*}}$.

To further illustrate Assumption 2, consider the model introduced in the appendix of  \citet{gunsilius2023distributional}, which is given by 
\begin{align}
\widetilde{Y}_{jtm, N} = \alpha_t + \beta_t U_{j,m},\label{m.2}
\end{align}
where $\alpha_t$ and $\beta_{t}$ are unknown parameters, and $U_{j,m}$ are independent and identically distributed draws from the unobservable distribution $F_{U_j}$, i.e., $U_{j,m} = F^{-1}_{U_j}(V_m)$. Similarly, we can express $\widehat{\widetilde{Y}}_{itm,N}$ in the same functional form as (\ref{m.2}): $\widehat{\widetilde{Y}}_{itm, N} = \alpha_t + \beta_t U_{j,m^{*}}$.

The appendix of \citet{gunsilius2023distributional} provides an identification motivation suggesting that stable identification of the counterfactual distribution naturally points to an affine (scaled-isometry) relationship between pre- and post-treatment distributional objects (see the discussion and figure therein). 
Model (\ref{m.2}) provides a concrete affine structure at the quantile level. Moreover, (\ref{m.2}) exhibits a one-factor structure at the quantile level and is closely aligned with the factor-type illustration in (\ref{m.1}).

We now show that Assumption 2 can be derived from more general assumptions as follows. 
The proof of how these subassumptions jointly imply Assumption 2 is provided in Appendix D.


\noindent\textbf{Assumption 2.1$^\prime$.}~~{\it There exists a constant $C_3$ such that $\mathbb{E}(U_{j, m^{*}}) < C_3$ for $j \in \{1,\ldots, J+1\}$ and $m\in\{1, 2, \ldots, M\}$.}

\noindent\textbf{Assumption 2.2$^\prime$.}~~{\it $T_1^{-1} \sum_{t_1\in \mathcal{T}_1} (\beta^2_t - \beta^2_{t_1}) = 0$ for $t \in \mathcal{T}_0$.}



\noindent\textbf{Assumption 2.3$^\prime$.}~~{\it $M^{-1} \sum_{m=1}^M \mathbb{E} (U_{i,m^*} U_{j,m^*}) - M^{-1} \sum_{m=1}^M \mathbb{E} (U_{i,m} U_{j,m}) = O (n^{-1/2})$ for any $i,j \in \{1,\ldots, J+1\}$.}


Assumption 2.1$^\prime$ imposes a uniform upper bound on the expectation of the latent variables $U_{j,m^*}$.
Assumption 2.2$^\prime$ requires that the squared coefficient $\beta^2_t$ in any pre-treatment period $t$ equals the average of $\beta^2_{t_1}$ over all post-treatment periods $t_1$. Notably, this condition is trivially satisfied when the coefficients are time-invariant, i.e., $\beta_t = \beta_{t_1}$ for $t \in \mathcal{T}_0 \cup \mathcal{T}_1$. 
We emphasize that Assumption 2.2$^\prime$ is introduced to facilitate the derivation of Assumption 2 under the model (\ref{m.2}), and thus serves a technical purpose. If the objective is solely to establish the asymptotic optimality of the proposed estimator, this condition can be relaxed. A detailed discussion of this relaxation is provided in Appendix D.
Assumption 2.3$^\prime$ requires that there be no substantially discrepancy between the second-moment structure of the latent variables $U_{i,m}$ and $U_{i,m^*}$.

For completeness, an alternative discussion of Assumption 2 under a dynamic panel quantile autoregression framework is provided in Appendix E.

Assumption 3 imposes constraints on the dependency of the potential outcomes $\widehat{\widetilde{Y}}_{itm, N}$ across quantile draws $m$. 
We illustrate the mildness of Assumption~3 in two scenarios.
First, conditional on the observed data $\{Y_{l,jt}\}$, the empirical quantile function $\wh{F}^{-1}_{Y_{jtn_j}}$ is entirely determined, so $\wh{\wt{Y}}_{itm,N} = \wh{F}^{-1}_{Y_{itn_i}}(V_m)$ is a deterministic transformation of $V_m$ alone. 
Thus, the mixing requirement of $\{\wh{\wt{Y}}_{itm,N}\}$ is reduced entirely to a condition of $\{V_m\}$. 
Assumption~3 holds trivially, when $V_m \overset{iid}{\sim} U[0,1]$ as in \citet{gunsilius2023distributional}; when $\{V_m\}$ is generated by a standard pseudo-random number generator, Assumption~3 also holds because the dependence decays rapidly by construction. 
Furthermore, when we consider the randomness of the observed data $\{Y_{l,jt}\}$, note that the randomness of $\wh{F}_{Y_{itn_i}}$ shrinks when $n_i\to\infty$ and thus the dependence of $\wh{\wt{Y}}_{itm,N} = \wh{F}^{-1}_{Y_{itn_i}}(V_m)$  across $m$ becomes asymptotically weak (typically of order $1/n_i$). 
Hence, in large samples the variables $\wh{\wt{Y}}_{itm,N}$ behave nearly as independent draws from ${F}^{-1}_{Y_{it}}(\cdot)$, for any fixed $i\in\{1,\cdots,J+1\}$ and $t\in\mathcal{T}_0\cup\mathcal{T}_1$; therefore, Assumption~3 can be satisfied.

We next illustrate how Assumptions 3 translate into more general assumptions under the two concrete models (\ref{m.1}) and (\ref{m.2}) introduced above.

\noindent\textbf{Assumption 3 under the quantile factor model (\ref{m.1}).}
In this case, Assumption 3 can be ensured by imposing a weak-dependence condition on the common factors across $m$:
\noindent\textbf{Assumption 3.1.}~~{\it For any $i \in\{1, \ldots, J+1\}$ and $t \in \mathcal{T}_0 \cup \mathcal{T}_1$, $\{\boldsymbol{f}_{t, m}\}_{m=1}^M$ is either $\alpha$-mixing with the mixing coefficient $\alpha = -r /(r-2)$ or $\phi$-mixing with the mixing coefficient $\phi = -r /(2 r-1)$ for $r \geq 2$.
}

Assumption 3.1 ensures that the weak dependence among the common factors $\boldsymbol{f}_{t,m}$ decays sufficiently fast, enabling the application of uniform laws of large numbers and central limit theorems.

\noindent\textbf{Assumption 3 under the simple linear model (\ref{m.2}) of \citet{gunsilius2023distributional}.}
In model (\ref{m.2}), the sequence $\{U_{j,m}: m=1,\ldots,M\}$ consists of independent and identically distributed draws from the unobservable distribution $F_{U_j}$. 
Hence, the required weak-dependence condition in Assumption 3 is automatically satisfied in this setting.

Assumption 4 consists of two parts. Assumption 4 (i) implies that the fourth moments of all $\widehat{\widetilde{Y}}_{itm, N}$ can be uniformly bounded. Assumption 4 (ii) concerns the difference between the quantiles of potential outcomes $Y_{jt, N}$ of the treated and control units, ensuring that these variances do not degenerate as $M$ increases.
We now illustrate how Assumptions 4 can be ensured under the same two concrete models (\ref{m.1}) and (\ref{m.2}).

\noindent\textbf{Assumption 4 under the quantile factor model (\ref{m.1}).}
Define $\mathbf{\Sigma}_{f,t} = \mathbb E(f_{t,m}f_{t,m}^\top)$ for $t\in \mathcal{T}_0 \cup \mathcal{T}_1$ and $m\in\{1, 2, \ldots, M\}$.
Assumption 4 can then be further specified as follows:

\noindent\textbf{Assumption 4.1.}

{\it (i) There exists a constant $C_f$ such that $\mathbb{E} \{\|\boldsymbol{f}_{t, m}\|^4\} \leq C_f < \infty$ for $m \in \{1, 2, \ldots, M\}$ and $t \in \mathcal{T}_0 \cup \mathcal{T}_1$.}

{\it (ii) There exist a constant $\kappa_f$ such that  $\lambda_{\min}(\mathbf{\Sigma}_{f,t})\geq \kappa_f>0$ for $t \in \mathcal{T}_0 \cup \mathcal{T}_1$ and $m\in\{1, 2, \ldots, M\}$.}

Assumption 4.1 (i) ensures that the fourth moments of the common factors are uniformly bounded, while Assumption 4.1 (ii) requires that the factor covariance matrix $\mathbf{\Sigma}_{f,t}$ is uniformly positive definite over time, preventing the latent factors from becoming degenerate.

\noindent\textbf{Assumption 4 under the simple linear model (\ref{m.2}) of \citet{gunsilius2023distributional}.}
Define $e_{U,m}^{(i)} = U_{i,m} - U_{1,m}$ for $i \in \{2,\ldots,J+1\}$ and $m \in \{1,\ldots,M\}$. 
Assumption 4 can be decomposed as the follows:

\noindent\textbf{Assumption 4.1$^\prime$.}

{\it (i) There exists a constant $C_U$ such that $\mathbb{E} \{U_{j,m}^4\} \leq C_U < \infty$ for $j \in\{1, \ldots, J+1\}$ and $m \in \{1, 2, \ldots, M\}$.}

{\it (ii) There exists a constant $C_4$ such that $\operatorname{var} (M^{-1/2} \sum_{m=1}^M e_{U,m}^{(i)} e_{U,m}^{(j)}) \geq C_4 > 0$ for $M$ sufficiently large and for any $i, j \in\{2, \ldots, J+1\}$.}

Assumption 4.1$^\prime$ (i) ensures that the fourth moments of all $U_{j,m}$ can be uniformly bounded. Assumption 4.1$^\prime$ (ii) concerns the difference between the $U_{j,m}$ of the treated and control units, ensuring that these variances do not degenerate as $M$ increases.




Assumption 5 imposes both lower and upper bounds on the variability of the quantiles of outcomes for each pre-treatment period $t$ of control units. This assumption ensures that the variation among the outcome quantiles of the control units is neither too small nor too large, and it plays a crucial role in establishing the convergence of the DSC weight.

\section{Simulation}
In this section, Monte Carlo simulations are conducted in both model-free and quantile factor model setups to verify Theorems 1 and 2. First, we examine the asymptotic optimality of the DSC estimator and subsequently verify the convergence of DSC weight.

\subsection{Simulation in a model-free setup}

\subsubsection{Simulation design}
We consider the following simulation design to validate the  theoretical results in Section 3.
For the time periods $t = 1, \ldots, T$, and $m = 1, 2, \ldots, M$, $\widetilde{Y}_{1 t m}$ are drawn from $ \chi^2(\mu_{1})$, where $\mu_{1}=2$, and $\widetilde{Y}_{j t m}$ are drawn independently from $\mathcal{N}(\mu_{j}, \sigma_{j}^{2})$ for $j = 2, \ldots, J + 1$, where $\mu_{j} \sim U(3, 10)$ and $\sigma_{j} = 3$ ($j$ is odd) or 2.5 ($j$ is even). 
To allow for dependence across the sampled ranks, we generate $\{V_m\}_{m=1}^M$ in dependent pairs. Specifically, we first draw $M/2$ ranks $V_k^{(1)}\sim U[0,1]$ and construct paired ranks
\[
V_k^{(2)}=
\begin{cases}
V_k^{(1)}+\delta, & \text{if } V_k^{(1)}<1/2,\\
V_k^{(1)}-\delta, & \text{if } V_k^{(1)}\ge 1/2,
\end{cases}
\]
where $\delta=0.01$, so that the resulting $M$ ranks $\{V_m\}_{m=1}^M$ exhibit dependence across $m$ and remain in $(0,1)$. This construction is one convenient way to introduce within-sample dependence across $m$; other dependence-generating schemes could be used as well.
As mentioned above, we set $j=1$ as the treated unit, and $j = 2, \ldots, J + 1$ are the control units.
We set $J\in \{20, 50\}$, $M\in \{50, 100, 200, 400\}$, the number of pre-treatment periods $T_0 = 10$ and the number of post-treatment periods $T_1 = 5$. The number of replications is $R = 1000$.

\subsubsection{Simulation results}

In order to investigate the asymptotic optimality of the DSC estimator, we need to know $\bar{R}_{T_1}(\mathbf{w})$.
One can compute $\bar{R}_{T_1}(\mathbf{w})$ as follows:
$$\bar{R}_{T_1}(\mathbf{w}) = \frac{1}{T_1} \sum_{t \in \mathcal{T}_1} \int_0^1 \left( \sum_{j=2}^{J+1} w_{jt} F_{Y_{j t}}^{-1}(q) - F_{Y_{1 t}}^{-1}(q) \right)^2 dq.$$

%

Since any weight $\lambda_{t}~(t\in \mathcal{T}_0)$ satisfying $\lambda_{t}\geq 0$ and $\sum_{t \in \mathcal{T}_0} \lambda_{t} = 1$ can be used, for the sake of simplicity, we use equal weights $\lambda_{t} = 1/ {T_0}$. The weights $\mathbf{w}_t$ in each pre-treatment period $t \in \mathcal{T}_0$ are estimated by equation (\ref{3.2})
and the optimal weight vector $\mathbf{w}_{T_1}^{\text {opt}}$ for a given $T_1$ is obtained by $\mathbf{w}_{T_1}^{\text {opt}} = {\arg \min }_{\mathbf{w} \in \mathcal{H}} \bar{R}_{T_1}(\mathbf{w})$.

Figure 1 plots the ratio $\bar{R}_{T_1}(\widehat{\mathbf{w}})/\inf_{\mathrm{w} \in \mathcal{H}} \bar{R}_{T_1}(\mathbf{w})$, under $J=20$ (solid line) and $J=50$ (dashed line), averaged over 1000 replications, as $M$ increases. The curves of the ratio under $J=20$ and $J=50$ both monotonically decrease toward 1 as $M$ increases. This observation indicates that the averaged 2-Wasserstein distance of post-treatment periods of the DSC estimators converges to the lowest possible averaged 2-Wasserstein distance of post-treatment periods as $M$ increases. This result aligns with the asymptotic optimality stated in Theorems 1.

To investigate the convergence of the DSC weight, Figure 2 plots vector norm of the difference between the $\widehat{\mathbf{w}}$ and $\mathbf{w}_{T_1}^{\text {opt}}$ under $J=20$ (solid line) and $J=50$ (dashed line), averaged over 1000 replications, as $M$ increases. We can find that no matter $J=20$ or 50, $\| \widehat{\mathbf{w}} - \mathbf{w}_{T_1}^{\text {opt}} \|$ is monotonically decreasing as $M$ increases, which agrees with the convergence result in Theorem 2. At the same time, comparing the values obtained under the different $J$, we find that $\widehat{\mathbf{w}}$ converges faster when $J = 20$ than $J = 50$, which again agrees with Theorem 2 that the convergence rate slows down when $J$ increases. 

\begin{figure}[htb]
\centering
\includegraphics[width=0.55\textwidth]{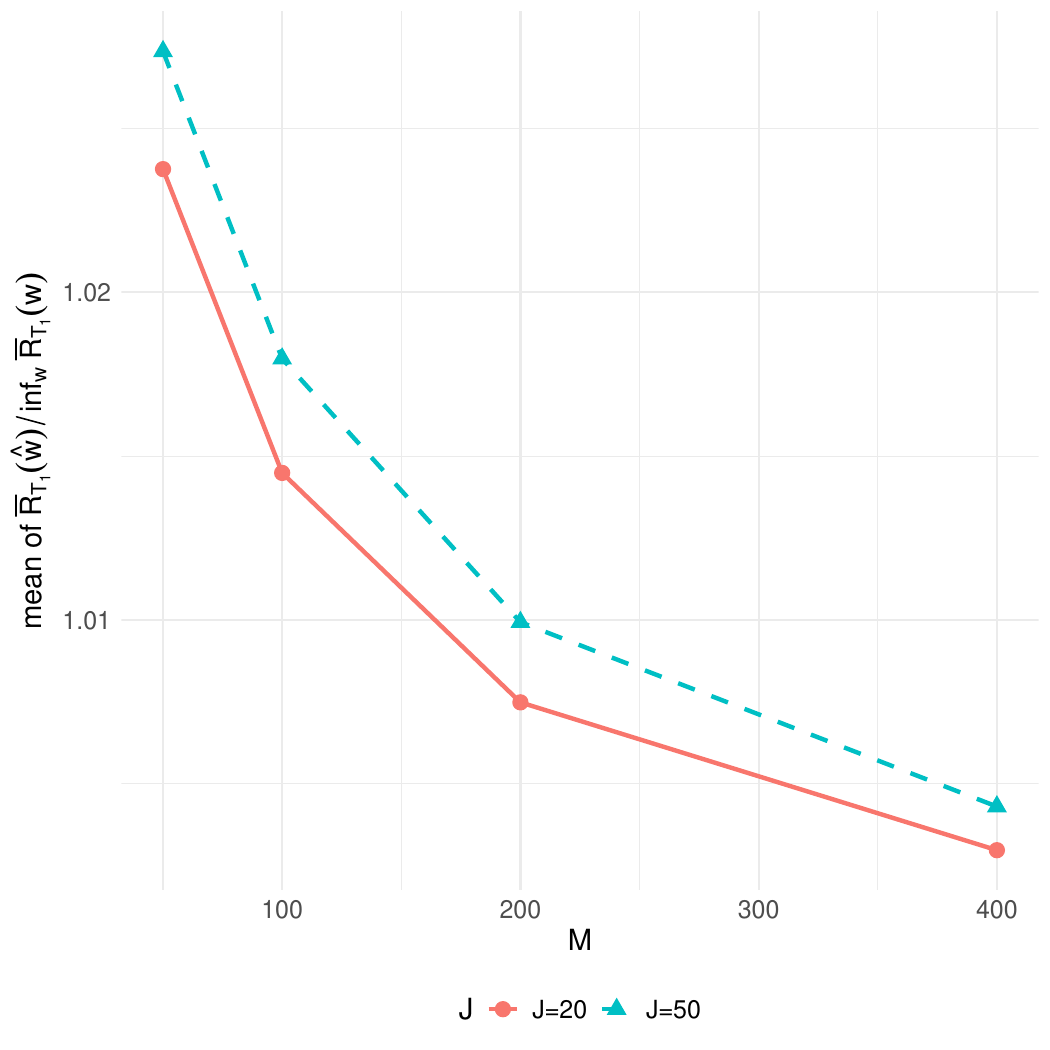}
\caption{Average of $\bar{R}_{T_1}(\widehat{\mathbf{w}})/\inf_{\mathrm{w} \in \mathcal{H}} \bar{R}_{T_1}(\mathbf{w})$ over 1000 replications}
\end{figure}

\begin{figure}[htb]
\centering
\includegraphics [width=0.55\textwidth]{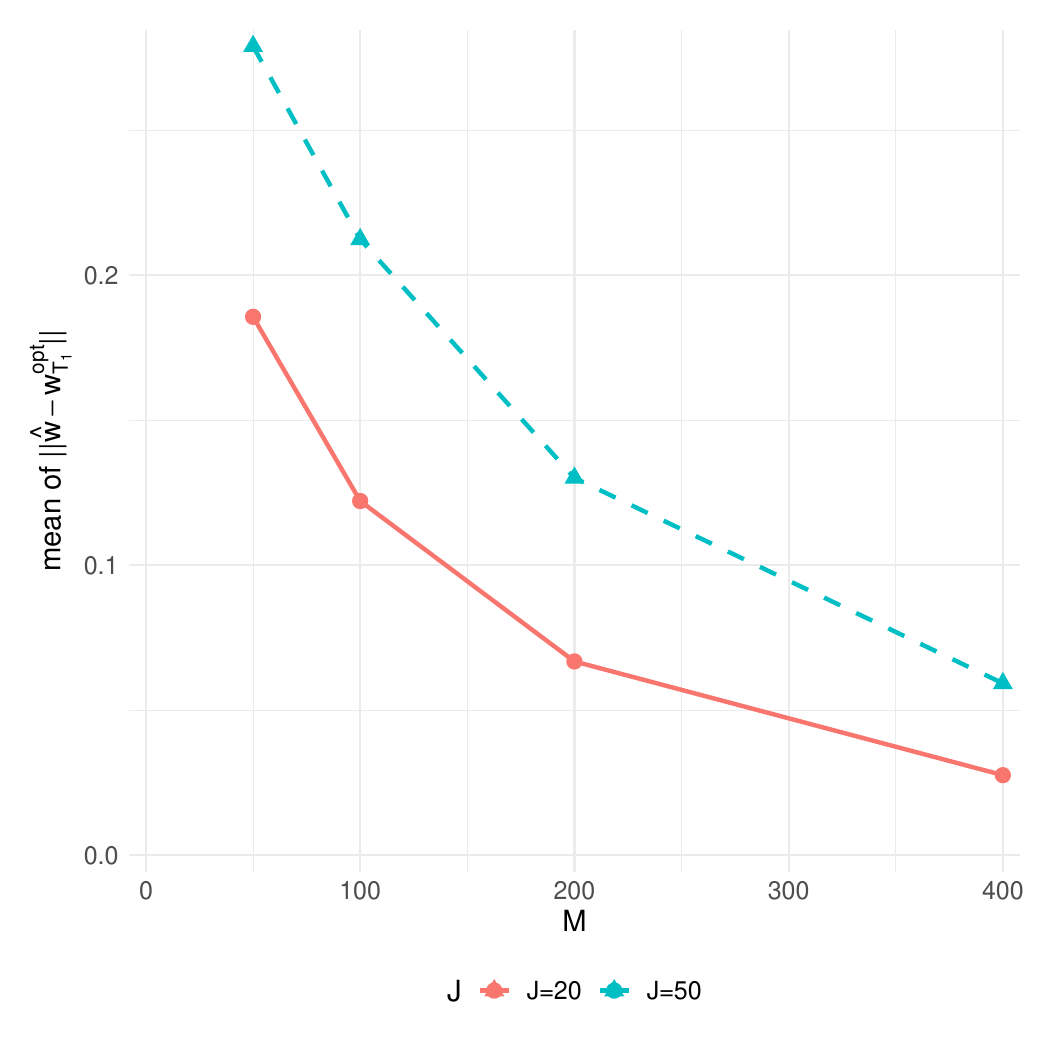}
\caption{Average of $\| \widehat{\mathbf{w}}-\mathbf{w}_{T_1}^{\text {opt}} \|$ over 1000 replications}
\end{figure}

\subsection{Simulation in a quantile factor model}

\subsubsection{Simulation design}
We generate the data from the following quantile factor structure:
$$\widetilde{Y}_{itm, N}  = \lambda_{1, i, m} f_{1, t, m} + \lambda_{2, i, m} f_{2, t, m},$$
where the common factors $f_{s, t, m}$, $s\in\{1,2\}$ are drawn independently from $\mathcal{N}(\mu_{t}, 3^{2})$ for $t = 1, \ldots, T$, and $\mu_{t}\sim \mathcal{N}(0, 1)$, and the factor loadings $\lambda_{s, i, m}$, $s\in\{1,2\}$ are drawn independently from $\mathcal{N}(\mu_{j}, \sigma_{j}^{2})$, where $\mu_{1}=2$, $\mu_{j} \sim U(2, 10)$ for $j = 2, \ldots, J + 1$, and $\sigma_{j} = 2.7$ ($j$ is odd) or 3 ($j$ is even). Similarly, we set $j=1$ as the treated unit, and $j = 2, \ldots, J + 1$ are the control units.
We set $J\in \{10, 20\}$, $M\in \{100, 200, 300, 400\}$, the number of pre-treatment periods $T_0 = 10$ and the number of post-treatment periods $T_1 = 5$. The number of replications is $R = 1000$.

\subsubsection{Simulation results}
As in the previous setup,
to investigate the asymptotic optimality and the convergence of the DSC weight, we present the results in Figure 3 and Figure 4.
Figure 3 plots the ratio $\bar{R}_{T_1}(\widehat{\mathbf{w}})/\inf _{\mathrm{w} \in \mathcal{H}} \bar{R}_{T_1}(\mathbf{w})$, under $J=10$ (solid line) and $J=20$ (dashed line), averaged over the 1000 replications, as $M$ increases. The curves of the ratio under $J=10$ and $J=20$ both monotonically decrease toward 1 as $M$ increases. This observation indicates that the averaged 2-Wasserstein distance of post-treatment periods of the DSC estimators converges to the lowest possible averaged 2-Wasserstein distance of post-treatment periods as $M$ increases. This result aligns with the asymptotic optimality stated in Theorems 1.

\begin{figure}[htb]
\centering
\includegraphics[width=0.55\textwidth]{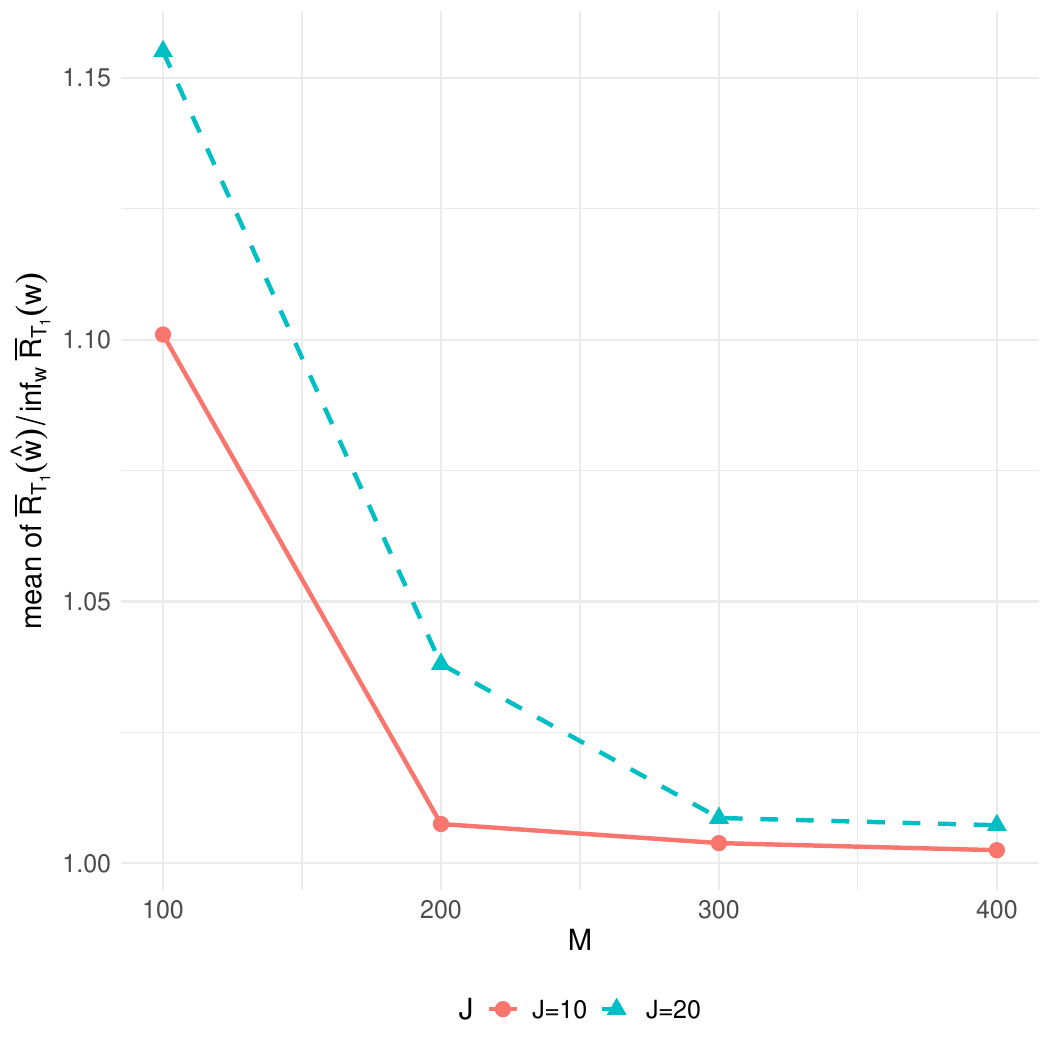}
\caption{Average of $\bar{R}_{T_1}(\widehat{\mathbf{w}})/\inf _{\mathrm{w} \in \mathcal{H}} \bar{R}_{T_1}(\mathbf{w})$ over 1000 replications}
\end{figure}

Figure 4 plots vector norm of the difference between the $\widehat{\mathbf{w}}$ and $\mathbf{w}_{T_1}^{\text {opt}}$ under $J=10$ (solid line) and $J=20$ (dashed line), averaged over the 1000 replications, as $M$ increases. We can find that no matter $J=10$ or 20, $\| \widehat{\mathbf{w}} - \mathbf{w}_{T_1}^{\text {opt}} \|$ is monotonically decreasing as $M$ increases, which agrees with the convergence result in Theorem 2. At the same time, comparing the values obtained under the different $J$, we find that $\widehat{\mathbf{w}}$ converges faster when $J = 10$ than $J = 20$, which again agrees with Theorem 2 that the convergence rate is slower when $J$ increases.

\begin{figure}[htb]
\centering
\includegraphics [width=0.55\textwidth]{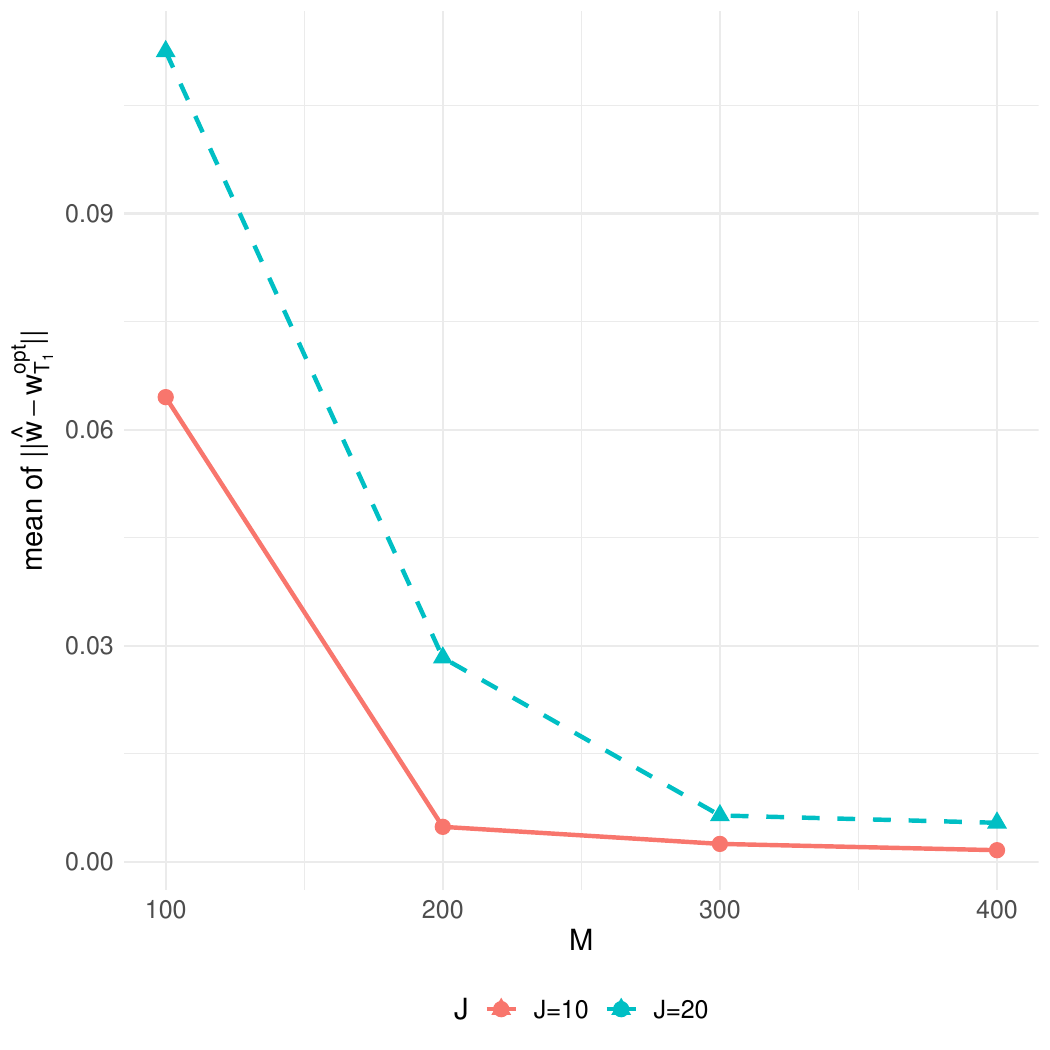}
\caption{Average of $\| \widehat{\mathbf{w}}-\mathbf{w}_{T_1}^{\text {opt}} \|$ over 1000 replications}
\end{figure}

\section{Conclusion}
In this paper, we investigate the asymptotic properties of the DSC estimator as $M\rightarrow \infty$.
We establish the asymptotic optimality of the DSC estimator, in the sense that
it achieves the lowest possible averaged 2-Wasserstein distance of post-treatment periods among all possible averaging estimators that are based on an average of quantiles of control units.
Furthermore, we show that the DSC weight converges to a limiting weight that minimizes the averaged 2-Wasserstein distance of post-treatment periods. At the same time, we quantify the rate of convergence, providing a better understanding of how the pre- and post-treatment fit, the number of control units and the number of draws $M$ influence the convergence rate.
Moreover, we present a natural extension of the DSC method in Appendix F, in which mixtures of quantile functions are replaced by mixtures of distribution functions. This alternative formulation may offer computational or interpretative advantages in certain applications, especially when working directly with estimated distribution functions.

\appendix

\section*{Appendix}
\addcontentsline{toc}{section}{Appendix}

\renewcommand{\theequation}{\thesection.\arabic{equation}}
\numberwithin{equation}{section}

\section{Proof and further discussion of Theorem 1}

\subsection{Proof of Theorem 1}
We prove (\ref{4.2}) in Theorem 1. When $T_0$ is finite, we can find a $t^{\prime} = \underset{t \in \mathcal{T}_0} {\operatorname{argmax}} \frac{\bar{R}_{T_1}(\widehat{\mathbf{w}}_t)} {\inf _{\mathbf{w} \in \mathcal{H}} \bar{R}_{T_1}(\mathbf{w})}$ (or $t^{\prime} = \underset{t \in \mathcal{T}_0} {\operatorname{argmax}} \bar{R}_{T_1}(\widehat{\mathbf{w}}_t) $), then, we have $\frac{\bar{R}_{T_1}(\widehat{\mathbf{w}}_{t^{\prime}})} {\inf _{\mathbf{w} \in \mathcal{H}} \bar{R}_{T_1}(\mathbf{w})} = \frac{\bar{R}_{T_1}(\widehat{\mathbf{w}}_{t^{\prime}})} {\inf _{\mathbf{w}_{t^{\prime}} \in \mathcal{H}} \bar{R}_{T_1}(\mathbf{w}_{t^{\prime}})}$. To prove (\ref{4.2}), it suffices to show that
\begin{align}
\frac{\bar{R}_{T_1}(\widehat{\mathbf{w}}_{t^{\prime}})}{\inf _{\mathbf{w}_{t^{\prime}} \in \mathcal{H}} \bar{R}_{T_1}(\mathbf{w}_{t^{\prime}})} \stackrel{p}{\rightarrow} 1. \label{B1}
\end{align}

We decompose $L_{t^{\prime}}(\mathbf{w}_{t^{\prime}})$ as
$$ L_{t^{\prime}}(\mathbf{w}_{t^{\prime}}) = \bar{R}_{T_1}(\mathbf{w}_{t^{\prime}}) + L_{t^{\prime}}(\mathbf{w}_{t^{\prime}}) - R_{t^{\prime}} (\mathbf{w}_{t^{\prime}}) + R_{t^{\prime}}(\mathbf{w}_{t^{\prime}}) - \bar{R}_{T_1}(\mathbf{w}_{t^{\prime}}). $$
By Lemma 1 in \citet{gao2019frequentist}, it suffices to show that
\begin{align}
\sup _{\mathbf{w}_{t^{\prime}} \in \mathcal{H}} \left| \frac{L_{t^{\prime}}(\mathbf{w}_{t^{\prime}}) - R_{t^{\prime}} (\mathbf{w}_{t^{\prime}})} {R_{t^{\prime}}(\mathbf{w}_{t^{\prime}})} \right| = o_p(1) \label{B2}
\end{align}
and
\begin{align}
\sup _{\mathbf{w}_{t^{\prime}} \in \mathcal{H}} \left| \frac{R_{t^{\prime}}(\mathbf{w}_{t^{\prime}}) - \bar{R}_{T_1}(\mathbf{w}_{t^{\prime}})} {R_{t^{\prime}}(\mathbf{w}_{t^{\prime}})} \right| = o(1). \label{B3}
\end{align}

We first verify (\ref{B2}). Note that
\begin{eqnarray}
&&\sup _{\mathbf{w}_{t^{\prime}} \in \mathcal{H}} \left| \frac{L_{t^{\prime}}(\mathbf{w}_{t^{\prime}}) - R_{t^{\prime}} (\mathbf{w}_{t^{\prime}})} {R_{t^{\prime}}(\mathbf{w}_{t^{\prime}})} \right| \notag\\
&\leq & \xi_{t^{\prime}}^{-1} \sup _{\mathbf{w}_{t^{\prime}} \in \mathcal{H}} \left| L_{t^{\prime}}(\mathbf{w}_{t^{\prime}}) - R_{t^{\prime}} (\mathbf{w}_{t^{\prime}}) \right| \notag\\
&=& \xi_{t^{\prime}}^{-1} \sup _{\mathbf{w}_{t^{\prime}} \in \mathcal{H}} \left| \frac{1}{M} \sum_{m=1}^M \left\{ \left( \sum_{j=2}^{J+1} w_{j t^{\prime}} \widehat{\widetilde{Y}}_{j t^{\prime} m} - \widehat{\widetilde{Y}}_{1 t^{\prime} m} \right)^2 - \mathbb{E} \left( \sum_{j=2}^{J+1} w_{j t^{\prime}} \widehat{\widetilde{Y}}_{j t^{\prime} m} - \widehat{\widetilde{Y}}_{1 t^{\prime} m} \right)^2 \right\} \right| \notag\\
&= & \xi_{t^{\prime}}^{-1} \sup _{\mathbf{w}_{t^{\prime}} \in \mathcal{H}} \left| \frac{1}{M} \sum_{m=1}^M \sum_{i=2}^{J+1} \sum_{j=2}^{J+1} w_{i t^{\prime}} w_{j t^{\prime}} \left\{ \left( \widehat{\widetilde{Y}}_{i t^{\prime} m} - \widehat{\widetilde{Y}}_{1 t^{\prime} m} \right) \left( \widehat{\widetilde{Y}}_{j t^{\prime} m} - \widehat{\widetilde{Y}}_{1 t^{\prime} m} \right) \right.\right. \notag\\
&&\left.\left. - \mathbb{E}\left( \widehat{\widetilde{Y}}_{i t^{\prime} m} - \widehat{\widetilde{Y}}_{1 t^{\prime} m} \right) \left( \widehat{\widetilde{Y}}_{j t^{\prime} m} - \widehat{\widetilde{Y}}_{1 t^{\prime} m} \right) \right\} \right| \notag\\
&\leq& \xi_{t^{\prime}}^{-1} \sup _{\mathbf{w}_{t^{\prime}} \in \mathcal{H}} \sum_{i=2}^{J+1} \sum_{j=2}^{J+1} \left| w_{i t^{\prime}} \right| \left| w_{j t^{\prime}} \right| \left| \frac{1}{M} \sum_{m=1}^M \left\{ \left( \widehat{\widetilde{Y}}_{i t^{\prime} m} - \widehat{\widetilde{Y}}_{1 t^{\prime} m} \right) \left( \widehat{\widetilde{Y}}_{j t^{\prime} m} - \widehat{\widetilde{Y}}_{1 t^{\prime} m} \right) \right.\right. \notag\\
&&\left.\left. - \mathbb{E}\left( \widehat{\widetilde{Y}}_{i t^{\prime} m} - \widehat{\widetilde{Y}}_{1 t^{\prime} m} \right) \left( \widehat{\widetilde{Y}}_{j t^{\prime} m} - \widehat{\widetilde{Y}}_{1 t^{\prime} m} \right) \right\} \right| \notag\\
&\leq & \xi_{t^{\prime}}^{-1} \sum_{i=2}^{J+1} \sum_{j=2}^{J+1} \left| \frac{1}{M} \sum_{m=1}^M \left\{ \left( \widehat{\widetilde{Y}}_{i t^{\prime} m} - \widehat{\widetilde{Y}}_{1 t^{\prime} m} \right) \left( \widehat{\widetilde{Y}}_{j t^{\prime} m} - \widehat{\widetilde{Y}}_{1 t^{\prime} m} \right) \right.\right. \notag\\
&&\left.\left. - \mathbb{E}\left( \widehat{\widetilde{Y}}_{i t^{\prime} m} - \widehat{\widetilde{Y}}_{1 t^{\prime} m} \right) \left( \widehat{\widetilde{Y}}_{j t^{\prime} m} - \widehat{\widetilde{Y}}_{1 t^{\prime} m} \right) \right\}  \right|  \notag\\
&=& \xi_{t^{\prime}}^{-1} M^{-1 / 2} \sum_{i=2}^{J+1} \sum_{j=2}^{J+1} \Psi_{M, t^{\prime}}(i, j), \label{B4}
\end{eqnarray}
where
\begin{eqnarray*}
\Psi_{M, t^{\prime}}(i, j) &=& \left|\frac{1}{\sqrt{M}} \sum_{m=1}^M \left\{ \left( \widehat{\widetilde{Y}}_{i t^{\prime} m} - \widehat{\widetilde{Y}}_{1 t^{\prime} m} \right) \left( \widehat{\widetilde{Y}}_{j t^{\prime} m} - \widehat{\widetilde{Y}}_{1 t^{\prime} m} \right) \right.\right.\\
&&\left.\left.- \mathbb{E}\left( \widehat{\widetilde{Y}}_{i t^{\prime} m} - \widehat{\widetilde{Y}}_{1 t^{\prime} m} \right) \left( \widehat{\widetilde{Y}}_{j t^{\prime} m} - \widehat{\widetilde{Y}}_{1 t^{\prime} m} \right) \right\} \right| \\
&=& \left| \frac{1}{\sqrt{M}} \sum_{m=1}^M \left\{e_{t^{\prime}, m, \widehat{\widetilde{Y}}_{N}}^{(i)} e_{t^{\prime}, m, \widehat{\widetilde{Y}}_{N}}^{(j)} - \mathbb{E} e_{t^{\prime}, m, \widehat{\widetilde{Y}}_{N}}^{(i)} e_{t^{\prime}, m, \widehat{\widetilde{Y}}_{N}}^{(j)} \right\} \right|.
\end{eqnarray*}

From Theorem 3.49 in \citet{white1984asymptotic} and Assumption 3, $e_{t^{\prime}, m, \widehat{\widetilde{Y}}_{N}}^{(i)} e_{t^{\prime}, m, \widehat{\widetilde{Y}}_{N}}^{(j)}$ for $i, j \in\{2, \ldots, J+1\}$ and $m \in \{1, 2, \ldots, M\}$ is either an $\alpha$-mixing sequence with the mixing coefficient $\alpha=-r /(r-2)$ or a $\phi$-mixing sequence with the mixing coefficient $\phi=-r /(2 r-1), r \geq 2$. Moreover, $\operatorname{var} ( e_{t^{\prime}, m, \widehat{\widetilde{Y}}_{N}}^{(i)} e_{t^{\prime}, m, \widehat{\widetilde{Y}}_{N}}^{(j)} )$ can be uniformly bounded, as a result of Assumption 4 (i). Using Theorem 5.20 in \citet{white1984asymptotic}, such properties of $e_{t^{\prime}, m, \widehat{\widetilde{Y}}_{N}}^{(i)} e_{t^{\prime}, m, \widehat{\widetilde{Y}}_{N}}^{(j)}$ together with Assumption 4 (ii) imply that
\begin{align}
\sum_{i=2}^{J+1} \sum_{j=2}^{J+1} \Psi_{M, t^{\prime}}(i, j) = O_p (J^2). \label{B5}
\end{align}
Combining (\ref{B4}), (\ref{B5}) and Assumption 1, we can obtain (\ref{B2}).

Then, we show that (\ref{B3}) holds. From Assumption 2, we have that
\begin{eqnarray*}
&& \sup _{\mathbf{w}_{t^{\prime}} \in \mathcal{H}} \left| \frac{R_{t^{\prime}}(\mathbf{w}_{t^{\prime}}) - \bar{R}_{T_1}(\mathbf{w}_{t^{\prime}})} {R_{t^{\prime}}(\mathbf{w}_{t^{\prime}})} \right| \\
&\leq & \xi_{t^{\prime}}^{-1} \sup _{\mathbf{w}_{t^{\prime}} \in \mathcal{H}} \left| R_{t^{\prime}}(\mathbf{w}_{t^{\prime}}) - \bar{R}_{T_1}(\mathbf{w}_{t^{\prime}}) \right| \\
&=& \xi_{t^{\prime}}^{-1} \left( O( n^{-1/2} J^2 ) + o(\xi_{t^{\prime}}) \right) \\
&=& O \left(\xi_{t^{\prime}}^{-1} M^{-1/2} J^2 \right) + o(1).
\end{eqnarray*}
This equation, combined with Assumption 1, leads to (\ref{B3}).
This completes the proof of Theorem 1.

\subsection{Further discussion of the result in Theorem 1}

This subsection provides a further discussion of the result in Theorem 1 by explicitly
separating its content into two steps.
The first step is a purely statistical result concerning asymptotic optimality with respect
to the pre-treatment risk.
The second step relies on an identification condition that links the pre-treatment and post-treatment risks and shows that the post-treatment optimality statement in Theorem 1 can be obtained from the pre-treatment optimality result together with an identification assumption.

\subsubsection*{Step (a): Pre-treatment asymptotic optimality}

We begin by formulating an asymptotic optimality result for the pre-treatment period.

\begin{theorem}
\label{thm:pre_opt}
Suppose that Assumptions 1, 3, and~4 hold and that $T_0$ is finite.
Then, for any fixed $t \in \mathcal T_0$,
\begin{equation}
\frac{R_t(\widehat{\mathbf{w}}_t)}{\inf_{\mathbf{w}_t \in \mathcal H} R_t(\mathbf{w}_t)}
\xrightarrow{p} 1.
\label{B6}
\end{equation}
\end{theorem}

\begin{proof}
We decompose $L_t(\mathbf{w}_t)$ as
$$ L_t(\mathbf{w}_t) = R_t(\mathbf{w}_t) + \left(L_t(\mathbf{w}_t) - R_t (\mathbf{w}_t)\right). $$
By Lemma 1 in \citet{gao2019frequentist}, to prove (\ref{B6}), it suffices to show that
\begin{align}
\sup_{w_t \in \mathcal H}
\frac{|L_t(\mathbf{w}_t) - R_t(\mathbf{w}_t)|}{R_t(\mathbf{w}_t)} = o_p(1).\label{B7}
\end{align}
By an argument identical to that used to establish \eqref{B2}, Assumptions 1, 3, and 4
imply \eqref{B7}, which in turn yields \eqref{B6}.
\end{proof}

Theorem \ref{thm:pre_opt} is a purely statistical result.
It uses only pre-treatment data and assumptions that control the sampling behavior of the
empirical risk, and it does not involve any information from the post-treatment period.

\subsubsection*{Step (b): From pre-treatment to post-treatment optimality}

We now show how the pre-treatment asymptotic optimality result in Theorem~\ref{thm:pre_opt} implies, and is implied by, the post-treatment optimality statement
in Theorem 1 once the identification condition in Assumption 2 is imposed.

\begin{proposition}
\label{prop:equiv}
Fix $t \in \mathcal T_0$ and suppose that Assumption 2 holds.
Then, under Theorem \ref{thm:pre_opt}, we have
\[
\frac{\bar R_{T_1}(\widehat{\mathbf{w}}_t)}{\inf_{\mathbf{w}_t \in \mathcal H} \bar R_{T_1}(\mathbf{w}_t)}
\xrightarrow{p} 1.
\]
\end{proposition}

\begin{proof}
From Assumption 1-2, we have that
\begin{eqnarray*}
&& \sup_{\mathbf{w} \in \mathcal{H}}
\frac{|R_t(\mathbf{w}_t) - \bar R_{T_1}(\mathbf{w}_t)|}{R_t(\mathbf{w}_t)} \\
&\leq & \xi_{t}^{-1} \sup_{\mathbf{w} \in \mathcal{H}} \left| R_{t}(\mathbf{w}_{t}) - \bar{R}_{T_1}(\mathbf{w}_{t}) \right| \\
&=& o(1).
\end{eqnarray*}
As a consequence, there exists a sequence $\varepsilon_n \to 0$ such that, uniformly over
$\mathbf{w}_t \in \mathcal H$,
\[
(1-\varepsilon_n) R_t(\mathbf{w}_t) \le \bar R_{T_1}(\mathbf{w}_t) \le (1+\varepsilon_n) R_t(\mathbf{w}_t).
\]
In particular,
\[
(1-\varepsilon_n) R_t(\widehat{\mathbf{w}}_t) \le \bar R_{T_1}(\widehat{\mathbf{w}}_t) \le (1+\varepsilon_n) R_t(\widehat{\mathbf{w}}_t),
\]
and taking infima over $\mathbf{w}_t \in\mathcal H$ yields
\[
(1-\varepsilon_n)\inf_{\mathbf{w}_t\in\mathcal H} R_t(\mathbf{w}_t)
\le
\inf_{\mathbf{w}_t \in\mathcal H} \bar R_{T_1}(\mathbf{w}_t)
\le
(1+\varepsilon_n)\inf_{\mathbf{w}_t \in\mathcal H} R_t(\mathbf{w}_t).
\]
Therefore,
\[
\frac{1-\varepsilon_n}{1+\varepsilon_n} \cdot
\frac{R_t(\widehat{\mathbf{w}}_t)}{\inf_{\mathbf{w}_t \in\mathcal H} R_t(\mathbf{w}_t)}
\le
\frac{\bar R_{T_1}(\widehat{\mathbf{w}}_t)}{\inf_{\mathbf{w}_t \in\mathcal H} \bar R_{T_1}(\mathbf{w}_t)}
\le
\frac{1+\varepsilon_n}{1-\varepsilon_n} \cdot
\frac{R_t(\widehat{\mathbf{w}}_t)}{\inf_{\mathbf{w}_t \in\mathcal H} R_t(\mathbf{w}_t)}.
\]
Since $\varepsilon_n \to 0$, we have $\frac{1-\varepsilon_n}{1+\varepsilon_n}=1+o(1)$ and
$\frac{1+\varepsilon_n}{1-\varepsilon_n}=1+o(1)$, which implies
\[
\frac{\bar R_{T_1}(\widehat{\mathbf{w}}_t)}{\inf_{\mathbf{w}_t \in \mathcal H} \bar R_{T_1}(\mathbf{w}_t)}
=
\frac{R_t(\widehat{\mathbf{w}}_t)}{\inf_{\mathbf{w}_t \in \mathcal H} R_t(\mathbf{w}_t)} \cdot (1+o(1)),
\]
which shows that
\[
\frac{\bar R_{T_1}(\widehat{\mathbf{w}}_t)}{\inf_{\mathbf{w}_t \in \mathcal H} \bar R_{T_1}(\mathbf{w}_t)}
\xrightarrow{p} 1.
\]
\end{proof}

Finally, we explain how the above two steps recover the original statement of Theorem 1.
When $T_0$ is finite and the DSC weight is defined as
$\widehat{\mathbf{w}} = \sum_{t\leq T_{0}} \lambda_{t} \widehat{\mathbf{w}}_{t}$,
there exists $t^{\prime} \in \mathcal T_0$ such that
\[
\frac{\bar R_{T_1}(\widehat{\mathbf{w}})}{\inf_{\mathbf{w} \in \mathcal H} \bar R_{T_1}(\mathbf{w})}
\le
\frac{\bar R_{T_1}(\widehat{\mathbf{w}}_{t^{\prime}})}{\inf_{\mathbf{w}_t \in \mathcal H} \bar R_{T_1}(\mathbf{w}_t)}.
\]
Applying Proposition \ref{prop:equiv} to $t^{\prime}$ yields the asymptotic optimality result in~\eqref{4.2},
thereby recovering Theorem 1.

\section{Proof of Theorem 2}

Since $\widehat{\mathbf{w}} = \sum_{t\leq T_{0}} \lambda_{t} \widehat{\mathbf{w}}_{t}$ for $\lambda_{t}\geq 0$ and $\sum_{t\leq T_{0}} \lambda_{t} = 1$, to prove Theorem 2, it suffices to show that, for each $t \in \mathcal{T}_0$,
\begin{align}
\left\|\widehat{\mathbf{w}}_t-\mathbf{w}_{T_1}^{opt}\right\| = O_p \left(\xi_{t}^{1/2} + \bar{\xi}_{T_1}^{1/2} + M^{-1/4} J\right). \label{A01}
\end{align}

Denote $\tau_{t} = \xi_{t}^{1/2} + \bar{\xi}_{T_1}^{1/2} + M^{-1/4} J$ and let $\mathbf{u} \in \mathbb{R}^J$ such that $\|\mathbf{u}\| = C_\varepsilon$, where $C_\varepsilon$ is a large enough constant. According to \citet{fan2004nonconcave} and \citet{lu2015jackknife}, to prove (\ref{A01}), it suffices to show that, for any $\varepsilon > 0$, there exists a constant $C_{\varepsilon}$, such that
\begin{align}
\operatorname{Pr}\left\{\inf _{\|\mathbf{u}\| = C_\varepsilon, \left(\mathbf{w}_{T_1}^{\mathrm{opt}}+\tau_{t} \mathbf{u}\right) \in \mathcal{H}} L_{t}\left(\mathbf{w}_{T_1}^{\mathrm{opt}}+\tau_{t} \mathbf{u}\right) \geq L_{t}\left(\mathbf{w}_{T_1}^{\mathrm{opt}}\right) \right\} > 1-\varepsilon, \label{A02}
\end{align}
for a given $T_1$ and any sufficiently large $M$. Formula (\ref{A02}) implies that with probability tending to 1 there exists a (local) minimizer $\mathbf{w}_{t}^{\star}$ of $L_{t}(\mathbf{w})$ in the ball $\{\mathbf{w}_{T_1}^{\mathrm{opt}} + \tau_{t} \mathbf{u} :  (\mathbf{w}_{T_1}^{\mathrm{opt}} + \tau_{t} \mathbf{u}) \in \mathcal{H}, \|\mathbf{u}\| \leq C_\varepsilon \}$ such that $\|\mathbf{w}_{t}^{\star} - \mathbf{w}_{T_1}^{\text {opt }}\| = O_p (\tau_{t})$. From the convexity of $L_{t}(\mathbf{w})$ and $\mathcal{H}$, $\mathbf{w}_{t}^{\star}$ is also the unique global minimizer, i.e., $\mathbf{w}_{t}^{\star} = \widehat{\mathbf{w}}_t$.

Define $D_{t}(\mathbf{u}) = L_{t} (\mathbf{w}_{T_1}^{\text {opt }} + \tau_{t} \mathbf{u}) - L_{t}(\mathbf{w}_{T_1}^{\text {opt}})$. Then, we can decompose $D_{t}(\mathbf{u})$ as
\begin{align}
D_{t}(\mathbf{u}) & = \frac{1}{M} \left\|\widehat{\widetilde{\mathbb{Y}}}_t \left( \mathbf{w}_{T_1}^{\text {opt}} + \tau_{t} \mathbf{u} \right) - \widehat{\widetilde{\boldsymbol{Y}}}_{1t}  \right\|^2 - \frac{1}{M} \left\|\widehat{\widetilde{\mathbb{Y}}}_t \mathbf{w}_{T_1}^{\text {opt}} - \widehat{\widetilde{\boldsymbol{Y}}}_{1t} \right\|^2 \nonumber\\
& = \frac{2 \tau_{t}}{M} \left( \widehat{\widetilde{\mathbb{Y}}}_t \mathbf{w}_{T_1}^{\text {opt}} - \widehat{\widetilde{\boldsymbol{Y}}}_{1t} \right)^{\top} \widehat{\widetilde{\mathbb{Y}}}_t \mathbf{u} + \frac{\tau_{t}^2}{M} \left\|\widehat{\widetilde{\mathbb{Y}}}_t \mathbf{u}\right\|^2 \nonumber \\
& \equiv \Delta_{t,1} + \Delta_{t,2}, \label{A003}
\end{align}
where $\Delta_{t,1} = 2 \tau_{t} M^{-1} ( \widehat{\widetilde{\mathbb{Y}}}_t \mathbf{w}_{T_1}^{\text {opt}} - \widehat{\widetilde{\boldsymbol{Y}}}_{1t} )^{\top} \widehat{\widetilde{\mathbb{Y}}}_t \mathbf{u}$ and $\Delta_{t,2} = \tau_{t}^2 M^{-1} \|\widehat{\widetilde{\mathbb{Y}}}_t \mathbf{u}\|^2$. We show that $\Delta_{t,2}$ is the dominant term of $D_{t}(\mathbf{u})$ as follows.

First, we consider $\Delta_{t,2}$. From Assumption 5, we have that, with probability approaching 1,
\begin{align}
\kappa_1 \leq \lambda_{\min }\left(\frac{1}{M} \widehat{\widetilde{\mathbb{Y}}}_t^{\top} \widehat{\widetilde{\mathbb{Y}}}_t \right) \leq \lambda_{\max }\left( \frac{1}{M} \widehat{\widetilde{\mathbb{Y}}}_t^{\top} \widehat{\widetilde{\mathbb{Y}}}_t \right) \leq \kappa_2. \label{A03}
\end{align}
This further implies that, with probability approaching 1,
\begin{align}
\Delta_{t,2} \geq \frac{\tau_{t}^2}{M} \lambda_{\min } \left(\widehat{\widetilde{\mathbb{Y}}}_t^{\top} \widehat{\widetilde{\mathbb{Y}}}_t \right) \|\mathbf{u}\|^2 \geq \kappa_1 \tau_{t}^2\|\mathbf{u}\|^2. \label{A04}
\end{align}

Next, we consider $\Delta_{t,1}$. From Assumption 2 and $M=Cn$, we have that
\begin{align}
\sup _{\mathbf{w_t} \in \mathcal{H}} \left|R_{t}(\mathbf{w}_t) - \bar{R}_{T_1}(\mathbf{w}_t) \right|
= O(M^{-1/2} J^{2}) + o(\xi_{t}). \label{A05}
\end{align}
Equation (\ref{A05}) holds for $\mathbf{w}_{T_1}^{\text {opt }}$, and note that $\bar{R}_{T_1} (\mathbf{w}_{T_1}^{\text {opt}}) = \bar{\xi}_{T_1}$. Thus, we have that
\begin{align}
R_{t}(\mathbf{w}_{T_1}^{\mathrm{opt}}) - \bar{\xi}_{T_1} = O (M^{-1 / 2} J^2) + o (\xi_{t}). \label{A06}
\end{align}
Since $\mathbb{E} \| \widehat{\widetilde{\mathbb{Y}}}_t \mathbf{w}_{T_1}^{\text {opt}} - \widehat{\widetilde{\boldsymbol{Y}}}_{1t} \|^2 = M R_{t} (\mathbf{w}_{T_1}^{\text {opt}})$, using (\ref{A06}), we have that
\begin{align}
\Big\| \widehat{\widetilde{\mathbb{Y}}}_t \mathbf{w}_{T_1}^{\text {opt}} - \widehat{\widetilde{\boldsymbol{Y}}}_{1t} \Big\| = O_p (M^{1 / 2} \bar{\xi}_{T_1}^{1 / 2}) + O_p (M^{1 / 4} J) + o_p (M^{1 / 2} \xi_{t}^{1 / 2}). \label{A07}
\end{align}
Thus, we can obtain that
\begin{align}
\left|\Delta_{t,1}\right| & \leq \frac{2 \tau_{t}}{M} \Big\| \widehat{\widetilde{\mathbb{Y}}}_t \mathbf{w}_{T_1}^{\text {opt}} - \widehat{\widetilde{\boldsymbol{Y}}}_{1t} \Big\| \big\| \widehat{\widetilde{\mathbb{Y}}}_t \mathbf{u} \big\| \nonumber\\
& \leq \frac{2 \tau_{t}}{M} \Big\| \widehat{\widetilde{\mathbb{Y}}}_t \mathbf{w}_{T_1}^{\text {opt}} - \widehat{\widetilde{\boldsymbol{Y}}}_{1t} \Big\| \sqrt{\lambda_{\max } ( \widehat{\widetilde{\mathbb{Y}}}_t^{\top} \widehat{\widetilde{\mathbb{Y}}}_t )}\|\mathbf{u}\| \nonumber\\
& = O_p(\tau_{t} \bar{\xi}_{T_1}^{1/2} )\|\mathbf{u}\| + O_p(\tau_{t} M^{-1/4} J) \|\mathbf{u}\| + o_p (\tau_{t} \xi_{t}^{1 / 2})\|\mathbf{u}\|, \label{A08}
\end{align}
where the last equality follows from (\ref{A03}) and (\ref{A07}). By (\ref{A04}) and (\ref{A08}), and allowing $\|\mathbf{u}\|$ to be sufficiently large, $\Delta_{t,2}$ dominates $\Delta_{t,1}$ and is positive. This, in conjunction with (\ref{A003}), implies that $D_{t}(\mathbf{u}) \geq 0$ with probability approaching 1.
This establishes (\ref{A02}), and therefore completes the proof of Theorem 2.


\section{Proof of Assumption 2 under Assumptions 2.1-2.4}
Let $\mathbf{e}_{\boldsymbol{\lambda}, m}^{(i)} = \boldsymbol{\lambda}_{1, m} - \boldsymbol{\lambda}_{i, m}$ and $\mathbf{e}_{\boldsymbol{\lambda}, m^{*}}^{(i)} = \boldsymbol{\lambda}_{1, m^{*}} - \boldsymbol{\lambda}_{i, m^{*}}$ for $i \in\{1, \ldots, J+1\}$, $m \in \{1, 2, \ldots, M\}$ and $m^{*} \in \{1, 2, \ldots, M\}$.
For proving Assumption 2, we define
$$L^{0}_{T_1}(\mathbf{w}) = \frac{1}{T_1} \sum_{t \in \mathcal{T}_1} \frac{1}{M} \sum_{m = 1}^M \left(\sum_{j=2}^{J+1} w_{j} \widetilde{Y}_{j t m, N} - \widetilde{Y}_{1 t m, N} \right)^2,$$
which can be seen as an approximate version of $\bar{R}_{T_1}(\mathbf{w})$, and the corresponding risk function is defined as $R^{0}_{T_1}(\mathbf{w}) = \mathbb{E} L^{0}_{T_1}(\mathbf{w})$, and define
$$R^{*}_{T_1}(\mathbf{w}) = \frac{1}{T_1} \sum_{t \in \mathcal{T}_1} \frac{1}{M} \sum_{m = 1}^M \mathbb{E} \left(\sum_{j=2}^{J+1} w_{j} \widehat{\widetilde{Y}}_{j t m, N} - \widehat{\widetilde{Y}}_{1 t m, N} \right)^2.$$
Then, under the Assumption 2.1, we have that, for each $t \in \mathcal{T}_0$,
\begin{eqnarray}
&&\sup _{\mathbf{w}_t \in \mathcal{H}} \left| R_{t}(\mathbf{w}_t) - \bar{R}_{T_1}(\mathbf{w}_t) \right| \nonumber\\
& \leq &\sup _{\mathbf{w}_t \in \mathcal{H}} \left|R_{t}(\mathbf{w}_t) - R^{*}_{T_1}(\mathbf{w}_t) \right| + \sup _{\mathbf{w}_t \in \mathcal{H}} \left|R^{*}_{T_1}(\mathbf{w}_t) - R^{0}_{T_1}(\mathbf{w}_t) \right| + \sup _{\mathbf{w}_t \in \mathcal{H}} \left|R^{0}_{T_1}(\mathbf{w}_t) - \bar{R}_{T_1}(\mathbf{w}_t) \right| \nonumber\\
& = &\sup _{\mathbf{w}_t \in \mathcal{H}} \left| \frac{1}{M} \sum_{m=1}^M \mathbb{E} \left( \sum_{j=2}^{J+1} w_{jt} \widehat{\widetilde{Y}}_{jtm} - \widehat{\widetilde{Y}}_{1tm} \right)^2 - \frac{1}{T_1} \sum_{t_1 \in \mathcal{T}_1} \frac{1}{M} \sum_{m=1}^M \mathbb{E} \left(\sum_{j=2}^{J+1} w_{jt} \widehat{\widetilde{Y}}_{j t_1 m} - \widehat{\widetilde{Y}}_{1 t_1 m, N} \right)^2 \right| \nonumber\\
&& + \sup _{\mathbf{w}_t \in \mathcal{H}} \left| \frac{1}{T_1} \sum_{t_1 \in \mathcal{T}_1} \frac{1}{M} \sum_{m=1}^M \mathbb{E} \left( \sum_{j=2}^{J+1} w_{jt} \widehat{\widetilde{Y}}_{j t_1 m} - \widehat{\widetilde{Y}}_{1 t_1 m, N} \right)^2 \right. \nonumber\\
&& \left. - \frac{1}{T_1} \sum_{t_1 \in \mathcal{T}_1} \frac{1}{M} \sum_{m=1}^M \mathbb{E} \left( \sum_{j=2}^{J+1} w_{jt} \widetilde{Y}_{j t_1 m} - \widetilde{Y}_{1 t_1 m, N} \right)^2 \right| \nonumber\\
&& + \sup _{\mathbf{w}_t \in \mathcal{H}} \left|R^{0}_{T_1}(\mathbf{w}_t) - \bar{R}_{T_1}(\mathbf{w}_t) \right| \nonumber\\
& = & \sup _{\mathbf{w}_t \in \mathcal{H}} \left| \frac{1}{M} \sum_{m=1}^M \mathbb{E} \left(\sum_{j=2}^{J+1} w_{jt} \boldsymbol{f}_{t, m^{*}}^{\top} \mathbf{e}_{\boldsymbol{\lambda}, m^{*}}^{(j)} \right)^2 - \frac{1}{T_1} \sum_{t_1 \in \mathcal{T}_1} \frac{1}{M} \sum_{m=1}^M \mathbb{E} \left(\sum_{j=2}^{J+1} w_{jt} \boldsymbol{f}_{t_1, m^{*}}^{\top} \mathbf{e}_{\boldsymbol{\lambda}, m^{*}}^{(j)} \right)^2 \right| \nonumber\\
&& + \sup _{\mathbf{w}_t \in \mathcal{H}} \left| \frac{1}{T_1} \sum_{t_1 \in \mathcal{T}_1} \frac{1}{M} \sum_{m=1}^M \mathbb{E} \left(\sum_{j=2}^{J+1} w_{jt} \boldsymbol{f}_{t_1, m^{*}}^{\top} \mathbf{e}_{\boldsymbol{\lambda}, m^{*}}^{(j)} \right)^2 \right. \nonumber\\
&& \left.- \frac{1}{T_1} \sum_{t_1 \in \mathcal{T}_1} \frac{1}{M} \sum_{m=1}^M \mathbb{E} \left(\sum_{j=2}^{J+1} w_{jt} \boldsymbol{f}_{t_1, m}^{\top} \mathbf{e}_{\boldsymbol{\lambda}, m}^{(j)} \right)^2 \right| \nonumber\\
&& + \sup _{\mathbf{w}_t \in \mathcal{H}} \left|R^{0}_{T_1}(\mathbf{w}_t) - \bar{R}_{T_1}(\mathbf{w}_t) \right| \nonumber\\
& \equiv &~\mathbb{I}_1 + \mathbb{I}_2 + \mathbb{I}_3. \label{4.01}
\end{eqnarray}

First, we consider $\mathbb{I}_1$. From Assumption 2.2, for each $t \in \mathcal{T}_0$, we have
$$\frac{1}{T_1} \sum_{t_1 \in \mathcal{T}_1} \mathbb{E}\left( \frac{1}{M} \sum_{m=1}^M \boldsymbol{f}_{t, m^{*}}^{\top} \boldsymbol{f}_{t, m^{*}} - \frac{1}{M} \sum_{m=1}^M \boldsymbol{f}_{t_1, m^{*}}^{\top} \boldsymbol{f}_{t_1, m^{*}} \right) = O (n^{-1/2}),$$
and the components of $\boldsymbol{\lambda}_{i, m^{*}}$ are bounded; hence,
\begin{eqnarray}
\mathbb{I}_1 & = & \sup _{\mathbf{w}_t \in \mathcal{H}} \left| \sum_{i=2}^{J+1} \sum_{j=2}^{J+1} w_{it} w_{jt} \left\{ \frac{1}{M} \sum_{m=1}^M \mathbb{E} \left[ \mathbb{E} \left( \boldsymbol{f}_{t, m^{*}}^{\top} \mathbf{e}_{\boldsymbol{\lambda}, m^{*}}^{(i)} \mathbf{e}_{\boldsymbol{\lambda}, m^{*}}^{(j) \top} \boldsymbol{f}_{t, m^{*}} | \boldsymbol{f}_{t, m^{*}} \right) \right] \right.\right. \nonumber\\
&& \left.\left. - \frac{1}{T_1} \sum_{t_1 \in \mathcal{T}_1} \frac{1}{M} \sum_{m=1}^M \mathbb{E} \left[ \mathbb{E} \left( \boldsymbol{f}_{t_1, m^{*}}^{\top} \mathbf{e}_{\boldsymbol{\lambda}, m^{*}}^{(i)} \mathbf{e}_{\boldsymbol{\lambda}, m^{*}}^{(j) \top} \boldsymbol{f}_{t_1, m^{*}} | \boldsymbol{f}_{t_1, m^{*}} \right) \right] \right\} \right| \nonumber\\
& \leq & \sup _{\mathbf{w}_t \in \mathcal{H}} \sum_{i=2}^{J+1} \sum_{j=2}^{J+1} \left| w_{it} \right| \left| w_{jt} \right| \left| \frac{1}{M} \sum_{m=1}^M \mathbb{E} \left(\boldsymbol{f}_{t, m^{*}}^{\top} \mathbf{e}_{\boldsymbol{\lambda}, m^{*}}^{(i)} \mathbf{e}_{\boldsymbol{\lambda}, m^{*}}^{(j) \top} \boldsymbol{f}_{t, m^{*}} \right) \right. \nonumber\\
&& \left. - \frac{1}{T_1} \sum_{t_1 \in \mathcal{T}_1} \frac{1}{M} \sum_{m=1}^M \mathbb{E} \left( \boldsymbol{f}_{t_1, m^{*}}^{\top} \mathbf{e}_{\boldsymbol{\lambda}, m^{*}}^{(i)} \mathbf{e}_{\boldsymbol{\lambda}, m^{*}}^{(j) \top} \boldsymbol{f}_{t_1, m^{*}} \right) \right| \nonumber\\
& \leq & \sum_{i=2}^{J+1} \sum_{j=2}^{J+1} \left| \frac{1}{M} \sum_{m=1}^M \mathbb{E} \left( \boldsymbol{f}_{t, m^{*}}^{\top} \mathbf{e}_{\boldsymbol{\lambda}, m^{*}}^{(i)} \mathbf{e}_{\boldsymbol{\lambda}, m^{*}}^{(j) \top} \boldsymbol{f}_{t, m^{*}} \right) \right. \nonumber\\
&& \left. - \frac{1}{T_1} \sum_{t_1 \in \mathcal{T}_1} \frac{1}{M} \sum_{m=1}^M \mathbb{E} \left(\boldsymbol{f}_{t_1, m^{*}}^{\top} \mathbf{e}_{\boldsymbol{\lambda}, m^{*}}^{(i)} \mathbf{e}_{\boldsymbol{\lambda}, m^{*}}^{(j) \top} \boldsymbol{f}_{t_1, m^{*}}\right) \right| \nonumber\\
& = & \sum_{i=2}^{J+1} \sum_{j=2}^{J+1} \left| \mathbb{E} \left\{ \operatorname{tr} \left(\frac{1}{M} \sum_{m=1}^M  \mathbf{e}_{\boldsymbol{\lambda}, m^{*}}^{(i)} \mathbf{e}_{\boldsymbol{\lambda}, m^{*}}^{(j) \top} \boldsymbol{f}_{t, m^{*}} \boldsymbol{f}_{t, m^{*}}^{\top} \right) \right.\right. \nonumber\\ \nonumber\\
&& \left.\left. - \operatorname{tr} \left( \frac{1}{T_1} \sum_{t_1 \in \mathcal{T}_1} \frac{1}{M} \sum_{m=1}^M \mathbf{e}_{\boldsymbol{\lambda}, m^{*}}^{(i)} \mathbf{e}_{\boldsymbol{\lambda}, m^{*}}^{(j) \top} \boldsymbol{f}_{t_1, m^{*}} \boldsymbol{f}_{t_1, m^{*}}^{\top} \right) \right\} \right| \nonumber\\
& = & \sum_{i=2}^{J+1} \sum_{j=2}^{J+1} \left| \mathbb{E} \left\{ \operatorname{tr} \left\{ \frac{1}{M} \sum_{m=1}^M \left( \boldsymbol{f}_{t, m^{*}} \boldsymbol{f}_{t, m^{*}}^{\top} - \frac{1}{T_1} \sum_{t_1 \in \mathcal{T}_1} \boldsymbol{f}_{t_1, m^{*}} \boldsymbol{f}_{t_1, m^{*}}^{\top} \right) \mathbf{e}_{\boldsymbol{\lambda}, m^{*}}^{(i)} \mathbf{e}_{\boldsymbol{\lambda}, m^{*}}^{(j) \top} \right\} \right\} \right| \nonumber\\
& = & O ( n^{-1/2} J^2 ). \label{4.02}
\end{eqnarray}
Here, the last equality is satisfied due to the fixed $F_{m}$. We then examine $\mathbb{I}_2$. From Assumptions 2.3 and 2.4, we have that
\begin{eqnarray}
\mathbb{I}_2 & = & \sup _{\mathbf{w}_t \in \mathcal{H}} \left| \sum_{i=2}^{J+1} \sum_{j=2}^{J+1} w_{it} w_{jt} \left\{\frac{1}{T_1} \sum_{t_1 \in \mathcal{T}_1} \frac{1}{M} \sum_{m=1}^M \mathbb{E} \left[ \mathbb{E} \left( \boldsymbol{f}_{t_1, m^{*}}^{\top} \mathbf{e}_{\boldsymbol{\lambda}, m^{*}}^{(i)} \mathbf{e}_{\boldsymbol{\lambda}, m^{*}}^{(j) \top} \boldsymbol{f}_{t_1, m^{*}} | \boldsymbol{f}_{t_1, m^{*}} \right) \right] \right.\right. \nonumber\\
&& \left.\left. - \frac{1}{T_1} \sum_{t_1 \in \mathcal{T}_1} \frac{1}{M} \sum_{m=1}^M \mathbb{E} \left[ \mathbb{E} \left( \boldsymbol{f}_{t_1, m}^{\top} \mathbf{e}_{\boldsymbol{\lambda}, m}^{(i)} \mathbf{e}_{\boldsymbol{\lambda}, m}^{(j) \top} \boldsymbol{f}_{t_1, m} | \boldsymbol{f}_{t_1, m} \right) \right] \right\} \right| \nonumber\\
& \leq & \sup _{\mathbf{w}_t \in \mathcal{H}} \sum_{i=2}^{J+1} \sum_{j=2}^{J+1} \left| w_{it} \right| \left| w_{jt} \right| \left| \frac{1}{T_1} \sum_{t_1 \in \mathcal{T}_1} \frac{1}{M} \sum_{m=1}^M \mathbb{E} \left(\boldsymbol{f}_{t_1, m^{*}}^{\top} \mathbf{e}_{\boldsymbol{\lambda}, m^{*}}^{(i)} \mathbf{e}_{\boldsymbol{\lambda}, m^{*}}^{(j) \top} \boldsymbol{f}_{t_1, m^{*}} \right) \right. \nonumber\\
&& \left. - \frac{1}{T_1} \sum_{t_1 \in \mathcal{T}_1} \frac{1}{M} \sum_{m=1}^M \mathbb{E} \left( \boldsymbol{f}_{t_1, m}^{\top} \mathbf{e}_{\boldsymbol{\lambda}, m}^{(i)} \mathbf{e}_{\boldsymbol{\lambda}, m}^{(j) \top} \boldsymbol{f}_{t_1, m} \right) \right| \nonumber\\
& \leq &\sum_{i=2}^{J+1} \sum_{j=2}^{J+1} \left|  \frac{1}{T_1} \sum_{t_1 \in \mathcal{T}_1} \frac{1}{M} \sum_{m=1}^M \mathbb{E} \left(\boldsymbol{f}_{t_1, m^{*}}^{\top} \mathbf{e}_{\boldsymbol{\lambda}, m^{*}}^{(i)} \mathbf{e}_{\boldsymbol{\lambda}, m^{*}}^{(j) \top} \boldsymbol{f}_{t_1, m^{*}} \right) \right.\nonumber\\
&&\left. - \frac{1}{T_1} \sum_{t_1 \in \mathcal{T}_1} \frac{1}{M} \sum_{m=1}^M \mathbb{E} \left( \boldsymbol{f}_{t_1, m}^{\top} \mathbf{e}_{\boldsymbol{\lambda}, m}^{(i)} \mathbf{e}_{\boldsymbol{\lambda}, m}^{(j) \top} \boldsymbol{f}_{t_1, m} \right) \right| \nonumber\\
& = & \sum_{i=2}^{J+1} \sum_{j=2}^{J+1} \left| \mathbb{E} \left\{ \operatorname{tr} \left( \frac{1}{T_1} \sum_{t_1 \in \mathcal{T}_1} \frac{1}{M} \sum_{m=1}^M \mathbf{e}_{\boldsymbol{\lambda}, m^{*}}^{(i)} \mathbf{e}_{\boldsymbol{\lambda}, m^{*}}^{(j) \top} \boldsymbol{f}_{t_1, m^{*}} \boldsymbol{f}_{t_1, m^{*}}^{\top} \right) \right. \right. \nonumber\\
&& \left.\left.- \operatorname{tr} \left( \frac{1}{T_1} \sum_{t_1 \in \mathcal{T}_1} \frac{1}{M} \sum_{m=1}^M \mathbf{e}_{\boldsymbol{\lambda}, m}^{(i)} \mathbf{e}_{\boldsymbol{\lambda}, m}^{(j) \top} \boldsymbol{f}_{t_1, m} \boldsymbol{f}_{t_1, m}^{\top} \right) \right\} \right| \nonumber\\
& = & \sum_{i=2}^{J+1} \sum_{j=2}^{J+1} \left| \mathbb{E} \left\{ \operatorname{tr} \left(\frac{1}{T_1} \sum_{t_1 \in \mathcal{T}_1} C_0 \left(\frac{1}{M} \sum_{m=1}^M \boldsymbol{f}_{t_1, m^{*}} \boldsymbol{f}_{t_1, m^{*}}^{\top} - \frac{1}{M} \sum_{m=1}^M \boldsymbol{f}_{t_1, m} \boldsymbol{f}_{t_1, m}^{\top} \right) \right) \right\} \right| \nonumber\\
& = & O ( n^{-1/2} J^2 ), \label{4.03}
\end{eqnarray}
where $C_0$ is a constant. Finally, We consider $\mathbb{I}_3$.
\begin{eqnarray}
\mathbb{I}_3 & = & \sup _{\mathbf{w}_t \in \mathcal{H}} \left|R^{0}_{T_1}(\mathbf{w}_t) - \bar{R}_{T_1}(\mathbf{w}_t) \right| \nonumber\\
& = & \frac{1}{T_1} \sum_{t_1 \in \mathcal{T}_1} \sup _{\mathbf{w}_{t} \in \mathcal{H}} \left| \mathbb{E} \left\{\frac{1}{M} \sum_{m=1}^M \left( \sum_{j=2}^{J+1} w_{jt} \widetilde{Y}_{j t_1 m} - \widetilde{Y}_{1 t_1 m, N} \right)^2 \right\} \right. \nonumber\\
&& \left. - \int_0^1 \left( \sum_{j=2}^{J+1} w_{jt} F_{Y_{j t_{1}}}^{-1}(q) - F_{Y_{1 t_{1}, N}}^{-1}(q) \right)^2 dq \right| \nonumber\\
& = & \frac{1}{T_1} \sum_{t_1 \in \mathcal{T}_1} \sup _{\mathbf{w}_{t} \in \mathcal{H}} \left| \frac{1}{M} \sum_{m=1}^M \mathbb{E} \left( \sum_{j=2}^{J+1} w_{jt} \widetilde{Y}_{j t_1 m} - \widetilde{Y}_{1 t_1 m, N} \right)^2 \right. \nonumber\\
&& \left. - \int_0^1 \left( \sum_{j=2}^{J+1} w_{jt} F_{Y_{j t_{1}}}^{-1}(q) - F_{Y_{1 t_{1}, N}}^{-1}(q) \right)^2 dq \right| \nonumber\\
& = & \frac{1}{T_1} \sum_{t_1 \in \mathcal{T}_1} \sup _{\mathbf{w}_{t} \in \mathcal{H}} \left| \frac{1}{M} \sum_{m=1}^M \int_0^1 \left( \sum_{j=2}^{J+1} w_{jt} F_{Y_{j t_{1}}}^{-1}(V_m) - F_{Y_{1 t_{1}, N}}^{-1}(V_m) \right)^2 dV_m \right. \nonumber\\
&& \left. - \int_0^1 \left( \sum_{j=2}^{J+1} w_{jt} F_{Y_{j t_{1}}}^{-1}(q) - F_{Y_{1 t_{1}, N}}^{-1}(q) \right)^2 dq \right| \nonumber\\
& = & \frac{1}{T_1} \sum_{t_1 \in \mathcal{T}_1} \sup _{\mathbf{w}_{t} \in \mathcal{H}} \left| \frac{1}{M} \sum_{m=1}^M \int_0^1 \left( \sum_{j=2}^{J+1} w_{jt} F_{Y_{j t_{1}}}^{-1}(q) - F_{Y_{1 t_{1}, N}}^{-1}(q) \right)^2 dq \right. \nonumber\\
&& \left. - \int_0^1 \left( \sum_{j=2}^{J+1} w_{jt} F_{Y_{j t_{1}}}^{-1}(q) - F_{Y_{1 t_{1}, N}}^{-1}(q) \right)^2 dq \right| \nonumber\\
& = & 0. \label{4.04}
\end{eqnarray}
Together with (\ref{4.01})-(\ref{4.04}), we achieve Assumption 2.

\section{Proof of Assumption 2 under Assumptions 2.1$^\prime$-2.3$^\prime$}
Let $e_{U, m}^{(i)} = U_{i,m} - U_{1,m}$ 
and $e_{U, m^{*}}^{(i)} = U_{i, m^{*}} - U_{1, m^{*}}$ for $i \in\{1, \ldots, J+1\}$ and $m \in \{1, 2, \ldots, M\}$.
Then, we have that, for each $t \in \mathcal{T}_0$,
\begin{eqnarray}
&&\sup _{\mathbf{w}_t \in \mathcal{H}} \left| R_{t}(\mathbf{w}_t) - \bar{R}_{T_1}(\mathbf{w}_t) \right| \nonumber\\
& \leq &\sup _{\mathbf{w}_t \in \mathcal{H}} \left|R_{t}(\mathbf{w}_t) - R^{*}_{T_1}(\mathbf{w}_t) \right| + \sup _{\mathbf{w}_t \in \mathcal{H}} \left|R^{*}_{T_1}(\mathbf{w}_t) - R^{0}_{T_1}(\mathbf{w}_t) \right| + \sup _{\mathbf{w}_t \in \mathcal{H}} \left|R^{0}_{T_1}(\mathbf{w}_t) - \bar{R}_{T_1}(\mathbf{w}_t) \right| \nonumber\\
& = &\sup _{\mathbf{w}_t \in \mathcal{H}} \left| \frac{1}{M} \sum_{m=1}^M \mathbb{E} \left( \sum_{j=2}^{J+1} w_{jt} \widehat{\widetilde{Y}}_{jtm} - \widehat{\widetilde{Y}}_{1tm} \right)^2 - \frac{1}{T_1} \sum_{t_1 \in \mathcal{T}_1} \frac{1}{M} \sum_{m=1}^M \mathbb{E} \left( \sum_{j=2}^{J+1} w_{jt} \widehat{\widetilde{Y}}_{j t_1 m} - \widehat{\widetilde{Y}}_{1 t_1 m, N} \right)^2 \right| \nonumber\\
&& + \sup _{\mathbf{w}_t \in \mathcal{H}} \left| \frac{1}{T_1} \sum_{t_1 \in \mathcal{T}_1} \frac{1}{M} \sum_{m=1}^M \mathbb{E} \left( \sum_{j=2}^{J+1} w_{jt} \widehat{\widetilde{Y}}_{j t_1 m} - \widehat{\widetilde{Y}}_{1 t_1 m, N} \right)^2 \right. \nonumber\\
&& \left. - \frac{1}{T_1} \sum_{t_1 \in \mathcal{T}_1} \frac{1}{M} \sum_{m=1}^M \mathbb{E} \left( \sum_{j=2}^{J+1} w_{jt} \widetilde{Y}_{j t_1 m} - \widetilde{Y}_{1 t_1 m, N} \right)^2 \right| \nonumber\\
&& + \sup _{\mathbf{w}_t \in \mathcal{H}} \left|R^{0}_{T_1}(\mathbf{w}_t) - \bar{R}_{T_1}(\mathbf{w}_t) \right| \nonumber\\
& = & \sup _{\mathbf{w}_t \in \mathcal{H}} \left| \frac{1}{M} \sum_{m=1}^M \mathbb{E} \left(\sum_{j=2}^{J+1} w_{jt} \beta_t {e}_{U, m^{*}}^{(j)} \right)^2 - \frac{1}{T_1} \sum_{t_1 \in \mathcal{T}_1} \frac{1}{M} \sum_{m=1}^M \mathbb{E} \left(\sum_{j=2}^{J+1} w_{jt} \beta_{t_1} e_{U, m^{*}}^{(j)} \right)^2 \right| \nonumber\\
&& + \sup _{\mathbf{w}_t \in \mathcal{H}} \left| \frac{1}{T_1} \sum_{t_1 \in \mathcal{T}_1} \frac{1}{M} \sum_{m=1}^M \mathbb{E} \left(\sum_{j=2}^{J+1} w_{jt} \beta_{t_1} e_{U, m^{*}}^{(j)} \right)^2 \right. \nonumber\\
&& \left.- \frac{1}{T_1} \sum_{t_1\in\mathcal{T}_1} \frac{1}{M} \sum_{m=1}^M \mathbb{E} \left(\sum_{j=2}^{J+1} w_{jt} \beta_{t_1} e_{U, m}^{(j)} \right)^2 \right| \nonumber\\
&& + \sup _{\mathbf{w}_t \in \mathcal{H}} \left|R^{0}_{T_1}(\mathbf{w}_t) - \bar{R}_{T_1}(\mathbf{w}_t) \right| \nonumber\\
& \equiv &~\mathbb{D}_1 + \mathbb{D}_2 + \mathbb{D}_3. \label{d.01}
\end{eqnarray}

First, we consider $\mathbb{D}_1$. From Assumptions 2.1$^\prime$ and 2.2$^\prime$, for each $t \in \mathcal{T}_0$, we have
\begin{eqnarray}
\mathbb{D}_1 & = & \sup _{\mathbf{w}_t \in \mathcal{H}} \left| \sum_{i=2}^{J+1} \sum_{j=2}^{J+1} w_{it} w_{jt} \left\{\frac{1}{M} \sum_{m=1}^M \mathbb{E} \left( \beta_t^2 e_{U, m^{*}}^{(i)} e_{U, m^{*}}^{(j)} \right) \right.\right. \nonumber\\
&& \left.\left. - \frac{1}{T_1} \sum_{t_1 \in \mathcal{T}_1} \frac{1}{M} \sum_{m=1}^M \mathbb{E} \left( \beta_{t_1}^2 e_{U, m^{*}}^{(i)} e_{U, m^{*}}^{(j)} \right) \right\} \right| \nonumber\\
& \leq & \sup _{\mathbf{w}_t \in \mathcal{H}} \sum_{i=2}^{J+1} \sum_{j=2}^{J+1} \left| w_{it} \right| \left| w_{jt} \right| \left| \frac{1}{M} \sum_{m=1}^M \mathbb{E} \left(\beta_t^2 e_{U, m^{*}}^{(i)} e_{U, m^{*}}^{(j)} \right) \right. \nonumber\\
&& \left. - \frac{1}{T_1} \sum_{t_1 \in \mathcal{T}_1} \frac{1}{M} \sum_{m=1}^M \mathbb{E} \left( \beta_{t_1}^2 e_{U, m^{*}}^{(i)} e_{U, m^{*}}^{(j)} \right) \right| \nonumber\\
& \leq & \sum_{i=2}^{J+1} \sum_{j=2}^{J+1} \left| \frac{1}{M} \sum_{m=1}^M \mathbb{E} \left( \beta_t^2 e_{U, m^{*}}^{(i)} e_{U, m^{*}}^{(j)} \right) - \frac{1}{T_1} \sum_{t_1 \in \mathcal{T}_1} \frac{1}{M} \sum_{m=1}^M \mathbb{E} \left( \beta_{t_1}^2 e_{U, m^{*}}^{(i)} e_{U, m^{*}}^{(j)} \right) \right| \nonumber\\
& = & \sum_{i=2}^{J+1} \sum_{j=2}^{J+1} \left| \mathbb{E} \left\{ (\beta_t^2 - \frac{1}{T_1} \sum_{t_1 \in \mathcal{T}_1} \beta_{t_1}^2) \frac{1}{M} \sum_{m=1}^M e_{U, m^{*}}^{(i)} e_{U, m^{*}}^{(j)} \right\} \right| \nonumber\\
& = & 0. \label{d.02}
\end{eqnarray}
We then examine $\mathbb{D}_2$. From Assumptions 2.1$^\prime$ and 2.3$^\prime$, we have that
\begin{eqnarray}
\mathbb{D}_2 & = & \sup _{\mathbf{w}_t \in \mathcal{H}} \left| \sum_{i=2}^{J+1} \sum_{j=2}^{J+1} w_{it} w_{jt} \left\{\frac{1}{T_1} \sum_{t_1 \in \mathcal{T}_1} \frac{1}{M} \sum_{m=1}^M \mathbb{E} \left( \beta_{t_1}^2 e_{U, m^{*}}^{(i)} e_{U, m^{*}}^{(j)} \right) \right.\right. \nonumber\\
&& \left.\left. - \frac{1}{T_1} \sum_{t_1 \in \mathcal{T}_1} \frac{1}{M} \sum_{m=1}^M \mathbb{E} \left( \beta_{t_1}^2 e_{U, m}^{(i)} e_{U, m}^{(j)} \right) \right\} \right| \nonumber\\
& \leq & \sup _{\mathbf{w}_t \in \mathcal{H}} \sum_{i=2}^{J+1} \sum_{j=2}^{J+1} \left| w_{it} \right| \left| w_{jt} \right| \left| \frac{1}{T_1} \sum_{t_1 \in \mathcal{T}_1} \frac{1}{M} \sum_{m=1}^M \mathbb{E} \left( \beta_{t_1}^2 e_{U, m^{*}}^{(i)} e_{U, m^{*}}^{(j)} \right) \right. \nonumber\\
&& \left. - \frac{1}{T_1} \sum_{t_1 \in \mathcal{T}_1} \frac{1}{M} \sum_{m=1}^M \mathbb{E} \left( \beta_{t_1}^2 e_{U, m}^{(i)} e_{U, m}^{(j)} \right) \right| \nonumber\\
& \leq &\sum_{i=2}^{J+1} \sum_{j=2}^{J+1} \left|  \frac{1}{T_1} \sum_{t_1 \in \mathcal{T}_1} \frac{1}{M} \sum_{m=1}^M \mathbb{E} \left(\beta_{t_1}^2 e_{U, m^{*}}^{(i)} e_{U, m^{*}}^{(j)} \right) \right.\nonumber\\
&&\left. - \frac{1}{T_1} \sum_{t_1 \in \mathcal{T}_1} \frac{1}{M} \sum_{m=1}^M \mathbb{E} \left( \beta_{t_1}^2 e_{U, m}^{(i)} e_{U, m}^{(j)} \right) \right| \nonumber\\
& = & \sum_{i=2}^{J+1} \sum_{j=2}^{J+1} \left| \frac{1}{T_1} \sum_{t_1 \in \mathcal{T}_1} \beta_{t_1}^2 \mathbb{E} \left(\frac{1}{M} \sum_{m=1}^M e_{U, m^{*}}^{(i)} e_{U, m^{*}}^{(j)}  - \frac{1}{M} \sum_{m=1}^M e_{U, m}^{(i)} e_{U, m}^{(j)} \right) \right| \nonumber\\
& = & O( n^{-1/2} J^2). \label{d.03}
\end{eqnarray}
Finally, We consider $\mathbb{D}_3$. From (\ref{4.04}), we have that
\begin{eqnarray}
\mathbb{D}_3 & = & \sup _{\mathbf{w}_t \in \mathcal{H}} \left|R^{0}_{T_1}(\mathbf{w}_t) - \bar{R}_{T_1}(\mathbf{w}_t) \right| \nonumber\\
& = & \frac{1}{T_1} \sum_{t_1 \in \mathcal{T}_1} \sup _{\mathbf{w}_{t} \in \mathcal{H}} \left| \mathbb{E} \left\{\frac{1}{M} \sum_{m=1}^M \left( \sum_{j=2}^{J+1} w_{jt} \widetilde{Y}_{j t_1 m} - \widetilde{Y}_{1 t_1 m, N} \right)^2 \right\} \right. \nonumber\\
&& \left. - \int_0^1 \left( \sum_{j=2}^{J+1} w_{jt} F_{Y_{j t_{1}}}^{-1}(q) - F_{Y_{1 t_{1}, N}}^{-1}(q) \right)^2 dq \right| \nonumber\\
& = & 0. \label{d.04}
\end{eqnarray}
Together with (\ref{d.01})-(\ref{d.04}), we achieve Assumption 2.

Next, we present a relaxed version of  Assumption 2.2$^\prime$ along with an additional regularity condition, and demonstrate that the asymptotic optimality result remains valid under these assumptions.

\noindent\textbf{Assumption D.1.}~~{\it
\it $T_1^{-1} \sum_{t_1\in \mathcal{T}_1} (\beta^2_t - \beta^2_{t_1}) = o(1)$ for $t \in \mathcal{T}_0$.
}

\noindent\textbf{Assumption D.2.}~~{\it
Let \(\epsilon_n\) be any sequence satisfying \(\epsilon_n=o(1)\). Assume that $\xi_t^{-1} J^2 \epsilon_n = o(1)$ for \(t\in\mathcal T_0\).
}

Assumption D.1 relaxes Assumption 2.2$^\prime$ by allowing for mild time variation in the parameters $\beta_t$. Assumption D.2 regulates the joint growth of the number of units \(J\) and \(\xi_t^{-1}\) relative to the convergence rate \(\epsilon_n\). This condition plays a role similar to Assumption 1 in the proof of asymptotic optimality.

Under Assumption 1, Assumptions D.1-D.2 and (\ref{d.01})-(\ref{d.04}), we have that, for each $t \in \mathcal{T}_0$, 
\begin{eqnarray*}
&& \sup _{\mathbf{w}_t\in \mathcal{H}} \left| \frac{R_{t}(\mathbf{w}_{t}) - \bar{R}_{T_1}(\mathbf{w}_{t})} {R_{t}(\mathbf{w}_{t})} \right| \\
&\leq & \xi_{t}^{-1} \sup _{\mathbf{w}_{t} \in \mathcal{H}} \left| R_{t}(\mathbf{w}_{t}) - \bar{R}_{T_1}(\mathbf{w}_{t}) \right| \\
& \leq & \xi_{t}^{-1} \left(\mathbb{D}_1 + \mathbb{D}_2 + \mathbb{D}_3\right) \\
&\leq & \xi_{t}^{-1} \sum_{i=2}^{J+1} \sum_{j=2}^{J+1} \left| \mathbb{E} \left\{ (\beta_t^2 - \frac{1}{T_1} \sum_{t_1 \in \mathcal{T}_1} \beta_{t_1}^2) \frac{1}{M} \sum_{m=1}^M e_{U, m^{*}}^{(i)} e_{U, m^{*}}^{(j)} \right\} \right| + O\left(\xi_{t}^{-1} n^{-1/2} J^2 \right)\\
&=& o(1).
\end{eqnarray*}
This equation leads to (\ref{B3}), which, together with the analysis of (\ref{B2}), completes the proof of asymptotic optimality.

\section{Discussion of Assumption 2 under a dynamic panel quantile autoregression model}

The assumption 2 considered in this paper can alternatively be motivated using more traditional nonlinear panel data models. In this appendix, we provide a discussion based on the dynamic panel quantile autoregression model introduced in \cite{galvao2011quantile} and \citet{arellano2016nonlinear}.
Specifically, suppose that the potential outcome $\widetilde{Y}_{itm, N}$ are generated according to
\begin{align}
\widetilde{Y}_{itm, N} = Q_{Y_{it, N}}(v_{m} | Y_{i,t-1, N}, \boldsymbol{X}_{it}) = \eta_{i}+\alpha_m Y_{i,t-1, N} + \boldsymbol{X}_{it}^{\prime}\boldsymbol{\beta}_m,\label{e.1}
\end{align}
where $\eta_i$ denotes the individual fixed effects, $Y_{i,t-1, N}$ is the lag of the outcome $Y_{it, N}$, and $\boldsymbol{X}_{it} = (X_{it,1}, \ldots, X_{it,p})^{\prime}$ is a $p$-vector of exogenous covariates. Similarly, we can express $\widehat{\widetilde{Y}}_{itm,N}$ as: $\widehat{\widetilde{Y}}_{itm, N} = \eta_{i}+\alpha_{m^{*}} Y_{i,t-1, N} + \boldsymbol{X}_{it}^{\prime}\boldsymbol{\beta}_{m^{*}}$.

We now show that Assumption 2 can be derived from more general assumptions as follows. 

\noindent\textbf{Assumption 2.1$^{\prime\prime}$.}~~{\it There exists a constant $C_e$ such that $\mathbb{E}|Y_{j, t-1}| < C_e$ and $\mathbb{E} \|\mathbf{X}_j\| < C_e$ for $j \in \{1,\ldots, J+1\}$ and $t\in\mathcal{T}_0 \cup \mathcal{T}_1$.}

\noindent\textbf{Assumption 2.2$^{\prime\prime}$.}

{\it (i) $M^{-1} \sum_{m=1}^M (\alpha^2_m - \alpha^2_{m^*}) = O(n^{-1/2})$ for $m\in\{1, 2, \ldots, M\}$.}

{\it (ii) $M^{-1} \sum_{m=1}^M (\boldsymbol{\beta}^2_m - \boldsymbol{\beta}^2_{m^*}) = O(n^{-1/2})$ for $m\in\{1, 2, \ldots, M\}$.}




\noindent\textbf{Assumption 2.3$^{\prime\prime}$.}~~{\it $\mathbb{E} (Y_{i,t-1,N} Y_{j,t-1,N} - T_1^{-1} \sum_{t_1 \in \mathcal{T}_{1}} Y_{i,t_1-1,N} Y_{j,t_1-1,N}) = 0$ for $t \in \mathcal{T}_0$ and $i,j \in \{1,\ldots, J+1\}$.}

\noindent\textbf{Assumption 2.4$^{\prime\prime}$.}~~{\it $\mathbb{E} (X_{it,l}X_{jt,l} - T_1^{-1} \sum_{t_1 \in \mathcal{T}_{1}} X_{it_1,l}X_{jt_1,l}) = 0$ for $t \in \mathcal{T}_0$, $l\in\{1, \ldots, p\}$ and $i,j \in \{1,\ldots, J+1\}$.}

Assumption 2.1$^{\prime\prime}$ imposes a uniform boundedness condition on the first moments of the lagged outcome and the covariates.
Assumption 2.2$^{\prime\prime}$ requires that the difference between the coefficients evaluated at the empirical quantile level $m$ and the corresponding population quantile level $m^*$ vanishes sufficiently fast as the sample size increases. 
Assumption 2.3$^{\prime\prime}$ imposes a stability condition on the second-moment structure of the lagged outcomes across the pre- and post-treatment periods. Specifically, it requires that cross-sectional second moments of $Y_{i,t-1}$ do not exhibit systematic shifts after treatment. This condition ensures that the relationship between the treated unit and the control units learned from the pre-treatment period remains informative for post-treatment counterfactual analysis.
Assumption 2.4$^{\prime\prime}$ is the analogue of Assumption 2.3$^{\prime\prime}$ for the covariates. It requires that the second-moment structure of the covariates remains stable across pre- and post-treatment periods.
We emphasize that Assumption 2.3$^{\prime\prime}$ and 2.4$^{\prime\prime}$ are introduced to facilitate the derivation of Assumption 2 under model (\ref{e.1}) and therefore serve a primarily technical purpose. If the goal is solely to establish the asymptotic optimality of the proposed estimator, these assumptions can be relaxed. A detailed discussion of such a relaxation is provided at the end of this appendix.

We now provide a derivation showing how these assumptions jointly imply Assumption 2. 
Let $\boldsymbol{\gamma}_{m} = (1, \alpha_m, \boldsymbol{\beta}^\prime_m)^{\prime}$, $\boldsymbol{\gamma}_{m^*} = (1, \alpha_m^*, \boldsymbol{\beta}^\prime_{m^*})^{\prime}$, $\boldsymbol{\mu}_{i,t} = (\eta_{i}, Y_{i,t-1, N}, \boldsymbol{X}_{it}^{\prime})^{\prime}$ and $\mathbf{e}_{\boldsymbol{\mu}}^{(i,t)} = \boldsymbol{\mu}_{i,t}-\boldsymbol{\mu}_{1,t}$
for $i \in\{1, \ldots, J+1\}$, $t \in \mathcal{T}_0 \cup \mathcal{T}_1$ and $m \in \{1, 2, \ldots, M\}$.
Then, we have that, for each $t \in \mathcal{T}_0$,
\begin{eqnarray}
&&\sup _{\mathbf{w}_t \in \mathcal{H}} \left| R_{t}(\mathbf{w}_t) - \bar{R}_{T_1}(\mathbf{w}_t) \right| \nonumber\\
& \leq &\sup _{\mathbf{w}_t \in \mathcal{H}} \left|R_{t}(\mathbf{w}_t) - R^{*}_{T_1}(\mathbf{w}_t) \right| + \sup _{\mathbf{w}_t \in \mathcal{H}} \left|R^{*}_{T_1}(\mathbf{w}_t) - R^{0}_{T_1}(\mathbf{w}_t) \right| + \sup _{\mathbf{w}_t \in \mathcal{H}} \left|R^{0}_{T_1}(\mathbf{w}_t) - \bar{R}_{T_1}(\mathbf{w}_t) \right| \nonumber\\
& = &\sup _{\mathbf{w}_t \in \mathcal{H}} \left| \frac{1}{M} \sum_{m=1}^M \mathbb{E} \left( \sum_{j=2}^{J+1} w_{jt} \widehat{\widetilde{Y}}_{jtm} - \widehat{\widetilde{Y}}_{1tm} \right)^2 \right. \nonumber\\
&& -\left. \frac{1}{T_1} \sum_{t_1 \in \mathcal{T}_1} \frac{1}{M} \sum_{m=1}^M \mathbb{E} \left( \sum_{j=2}^{J+1} w_{jt} \widehat{\widetilde{Y}}_{j t_1 m} - \widehat{\widetilde{Y}}_{1 t_1 m, N} \right)^2 \right| \nonumber\\
&& + \sup _{\mathbf{w}_t \in \mathcal{H}} \left| \frac{1}{T_1} \sum_{t_1 \in \mathcal{T}_1} \frac{1}{M} \sum_{m=1}^M \mathbb{E} \left( \sum_{j=2}^{J+1} w_{jt} \widehat{\widetilde{Y}}_{j t_1 m} - \widehat{\widetilde{Y}}_{1 t_1 m, N} \right)^2 \right. \nonumber\\
&& \left. - \frac{1}{T_1} \sum_{t_1 \in \mathcal{T}_1} \frac{1}{M} \sum_{m=1}^M \mathbb{E} \left( \sum_{j=2}^{J+1} w_{jt} \widetilde{Y}_{j t_1 m} - \widetilde{Y}_{1 t_1 m, N} \right)^2 \right| \nonumber\\
&& + \sup _{\mathbf{w}_t \in \mathcal{H}} \left|R^{0}_{T_1}(\mathbf{w}_t) - \bar{R}_{T_1}(\mathbf{w}_t) \right| \nonumber\\
& = & \sup _{\mathbf{w}_t \in \mathcal{H}} \left| \frac{1}{M} \sum_{m=1}^M \mathbb{E} \left(\sum_{j=2}^{J+1} w_{jt} (\boldsymbol{\gamma}_{m^*}^{\prime} \boldsymbol{\mu}_{j,t})-\boldsymbol{\gamma}_{m^*}^{\prime}\boldsymbol{\mu}_{1,t} \right)^2 \right. \nonumber\\
&& \left. - \frac{1}{T_1} \sum_{t_1 \in \mathcal{T}_1} \frac{1}{M} \sum_{m=1}^M \mathbb{E} \left(\sum_{j=2}^{J+1} w_{jt} (\boldsymbol{\gamma}_{m^*}^{\prime} \boldsymbol{\mu}_{j,t})-\boldsymbol{\gamma}_{m^*}^{\prime}\boldsymbol{\mu}_{1,t} \right)^2 \right| \nonumber\\
&& + \sup _{\mathbf{w}_t \in \mathcal{H}} \left| \frac{1}{T_1} \sum_{t_1 \in \mathcal{T}_1} \frac{1}{M} \sum_{m=1}^M \mathbb{E} \left(\sum_{j=2}^{J+1} w_{jt} (\boldsymbol{\gamma}_{m^*}^{\prime} \boldsymbol{\mu}_{j,t})-\boldsymbol{\gamma}_{m^*}^{\prime}\boldsymbol{\mu}_{1,t} \right)^2 \right. \nonumber\\
&& \left.- \frac{1}{T_1} \sum_{t\in\mathcal{T}_1} \frac{1}{M} \sum_{m=1}^M \mathbb{E} \left(\sum_{j=2}^{J+1} w_{jt} (\boldsymbol{\gamma}_{m}^{\prime} \boldsymbol{\mu}_{j,t})-\boldsymbol{\gamma}_{m}^{\prime}\boldsymbol{\mu}_{1,t} \right)^2 \right| \nonumber\\
&& + \sup _{\mathbf{w}_t \in \mathcal{H}} \left|R^{0}_{T_1}(\mathbf{w}_t) - \bar{R}_{T_1}(\mathbf{w}_t) \right| \nonumber\\
& = & \sup _{\mathbf{w}_t \in \mathcal{H}} \left| \frac{1}{M} \sum_{m=1}^M \mathbb{E} \left(\boldsymbol{\gamma}_{m^*}^{\prime} \mathbf{e}_{\boldsymbol{\mu},\mathbf{w}_t}^{(j,t)} \right)^2 - \frac{1}{T_1} \sum_{t_1 \in \mathcal{T}_1} \frac{1}{M} \sum_{m=1}^M \mathbb{E} \left(  \boldsymbol{\gamma}_{m^*}^{\prime} \mathbf{e}_{\boldsymbol{\mu},\mathbf{w}_t}^{(j,t)} \right)^2 \right| \nonumber\\
&& + \sup _{\mathbf{w}_t \in \mathcal{H}} \left| \frac{1}{T_1} \sum_{t_1 \in \mathcal{T}_1} \frac{1}{M} \sum_{m=1}^M \mathbb{E} \left(\sum_{j=2}^{J+1} w_{jt} \boldsymbol{\gamma}_{m^*}^{\prime} \mathbf{e}_{\boldsymbol{\mu}}^{(j,t)} \right)^2 - \frac{1}{T_1} \sum_{t\in\mathcal{T}_1} \frac{1}{M} \sum_{m=1}^M \mathbb{E} \left(\sum_{j=2}^{J+1} w_{jt} \boldsymbol{\gamma}_{m}^{\prime} \mathbf{e}_{\boldsymbol{\mu}}^{(j,t)} \right)^2 \right| \nonumber\\
&& + \sup _{\mathbf{w}_t \in \mathcal{H}} \left|R^{0}_{T_1}(\mathbf{w}_t) - \bar{R}_{T_1}(\mathbf{w}_t) \right| \nonumber\\
& \equiv &~\mathbb{E}_1 + \mathbb{E}_2 + \mathbb{E}_3. \label{e.2}
\end{eqnarray}

First, we consider $\mathbb{E}_1$. From Assumptions 2.3$^{\prime\prime}$ and 2.4$^{\prime\prime}$, we have

\begin{eqnarray}
\mathbb{E}_1 & = & \sup_{\mathbf{w}_t \in \mathcal{H}} \left| \sum_{i=2}^{J+1} \sum_{j=2}^{J+1} w_{it} w_{jt} \left\{ \frac{1}{M} \sum_{m=1}^M \mathbb{E} \left( \boldsymbol{\gamma}_{m^*}^{\prime} \mathbf{e}_{\boldsymbol{\mu}}^{(i,t)}  \mathbf{e}_{\boldsymbol{\mu}}^{(j,t) \prime} \boldsymbol{\gamma}_{m^*} \right) \right.\right. \nonumber\\ 
&& \left.\left.- \frac{1}{T_1} \sum_{t_1 \in \mathcal{T}_1} \frac{1}{M} \sum_{m=1}^M \mathbb{E} \left( \boldsymbol{\gamma}_{m^*}^{\prime} \mathbf{e}_{\boldsymbol{\mu}}^{(i,t)}  \mathbf{e}_{\boldsymbol{\mu}}^{(j,t) \prime} \boldsymbol{\gamma}_{m^*} \right) \right\} \right| \nonumber\\
& \leq & \sup _{\mathbf{w}_t \in \mathcal{H}} \sum_{i=2}^{J+1} \sum_{j=2}^{J+1} | w_{it}| |w_{jt}| \left| \left\{ \frac{1}{M} \sum_{m=1}^M \mathbb{E} \left( \boldsymbol{\gamma}_{m^*}^{\prime} \mathbf{e}_{\boldsymbol{\mu}}^{(i,t)}  \mathbf{e}_{\boldsymbol{\mu}}^{(j,t) \prime} \boldsymbol{\gamma}_{m^*} \right) \right.\right. \nonumber\\ 
&& \left.\left. - \frac{1}{T_1} \sum_{t_1 \in \mathcal{T}_1} \frac{1}{M} \sum_{m=1}^M \mathbb{E} \left(  \boldsymbol{\gamma}_{m^*}^{\prime} \mathbf{e}_{\boldsymbol{\mu}}^{(i,t)}  \mathbf{e}_{\boldsymbol{\mu}}^{(j,t) \prime} \boldsymbol{\gamma}_{m^*} \right) \right\} \right| \nonumber\\
& \leq & \sum_{i=2}^{J+1} \sum_{j=2}^{J+1} \left| \left\{ \frac{1}{M} \sum_{m=1}^M \mathbb{E} \left( \boldsymbol{\gamma}_{m^*}^{\prime} \mathbf{e}_{\boldsymbol{\mu}}^{(i,t)}  \mathbf{e}_{\boldsymbol{\mu}}^{(j,t) \prime} \boldsymbol{\gamma}_{m^*} \right) - \frac{1}{T_1} \sum_{t_1 \in \mathcal{T}_1} \frac{1}{M} \sum_{m=1}^M \mathbb{E} \left(  \boldsymbol{\gamma}_{m^*}^{\prime} \mathbf{e}_{\boldsymbol{\mu}}^{(i,t)}  \mathbf{e}_{\boldsymbol{\mu}}^{(j,t) \prime} \boldsymbol{\gamma}_{m^*} \right) \right\} \right| \nonumber\\ 
& = & \sum_{i=2}^{J+1} \sum_{j=2}^{J+1} \left| \mathbb{E} \left\{ \operatorname{tr} \left(\frac{1}{M} \sum_{m=1}^M \mathbf{e}_{\boldsymbol{\mu}}^{(i,t)} \mathbf{e}_{\boldsymbol{\mu}}^{(j,t) \prime} \boldsymbol{\gamma}_{m^*} \boldsymbol{\gamma}_{m^*}^{\prime} \right) \right.\right. \nonumber\\ 
&& \left.\left. - \operatorname{tr} \left(\frac{1}{T_1} \sum_{t_1 \in \mathcal{T}_1} \frac{1}{M} \sum_{m=1}^M \mathbf{e}_{\boldsymbol{\mu}}^{(i,t)} \mathbf{e}_{\boldsymbol{\mu}}^{(j,t) \prime} \boldsymbol{\gamma}_{m^*} \boldsymbol{\gamma}_{m^*}^{\prime} \right) \right\} \right| \nonumber\\
& = & \sum_{i=2}^{J+1} \sum_{j=2}^{J+1} \left| \mathbb{E} \left\{ \operatorname{tr} \left( \frac{1}{M} \sum_{m=1}^M (\mathbf{e}_{\boldsymbol{\mu}}^{(i,t)} \mathbf{e}_{\boldsymbol{\mu}}^{(j,t) \prime} - \mathbf{e}_{\boldsymbol{\mu}}^{(i,t_1)} \mathbf{e}_{\boldsymbol{\mu}}^{(j,t_1) \prime}) \boldsymbol{\gamma}_{m^*} \boldsymbol{\gamma}_{m^*}^{\prime} \right) \right\} \right| \nonumber\\
& = & 0. \label{e.3}
\end{eqnarray}
We then examine $\mathbb{E}_2$. From Assumptions 2.1$^{\prime\prime}$ and 2.2$^{\prime\prime}$, we have

\begin{eqnarray}
\mathbb{E}_2 & = & \sup _{\mathbf{w}_t \in \mathcal{H}} \left| \sum_{i=2}^{J+1} \sum_{j=2}^{J+1} w_{it} w_{jt} \left\{\frac{1}{T_1} \sum_{t_1 \in \mathcal{T}_1} \frac{1}{M} \sum_{m=1}^M \mathbb{E} \left( \boldsymbol{\gamma}_{m^*}^{\prime} \mathbf{e}_{\boldsymbol{\mu}}^{(i,t)}  \mathbf{e}_{\boldsymbol{\mu}}^{(j,t) \prime} \boldsymbol{\gamma}_{m^*} \right) \right.\right. \nonumber\\
&& \left.\left. - \frac{1}{T_1} \sum_{t_1 \in \mathcal{T}_1} \frac{1}{M} \sum_{m=1}^M \mathbb{E} \left( \boldsymbol{\gamma}_{m}^{\prime} \mathbf{e}_{\boldsymbol{\mu}}^{(i,t)}  \mathbf{e}_{\boldsymbol{\mu}}^{(j,t) \prime} \boldsymbol{\gamma}_{m} \right) \right\} \right| \nonumber\\
& \leq & \sup _{\mathbf{w}_t \in \mathcal{H}} \sum_{i=2}^{J+1} \sum_{j=2}^{J+1} |w_{it}| | w_{jt} | \left| \frac{1}{T_1} \sum_{t_1 \in \mathcal{T}_1} \frac{1}{M} \sum_{m=1}^M \mathbb{E} \left( \boldsymbol{\gamma}_{m^*}^{\prime} \mathbf{e}_{\boldsymbol{\mu}}^{(i,t)}  \mathbf{e}_{\boldsymbol{\mu}}^{(j,t) \prime} \boldsymbol{\gamma}_{m^*} \right) \right. \nonumber\\
&& \left. - \frac{1}{T_1} \sum_{t_1 \in \mathcal{T}_1} \frac{1}{M} \sum_{m=1}^M \mathbb{E} \left( \boldsymbol{\gamma}_{m}^{\prime} \mathbf{e}_{\boldsymbol{\mu}}^{(i,t)}  \mathbf{e}_{\boldsymbol{\mu}}^{(j,t) \prime} \boldsymbol{\gamma}_{m} \right) \right| \nonumber\\
& \leq &\sum_{i=2}^{J+1} \sum_{j=2}^{J+1} \left| \frac{1}{T_1} \sum_{t_1 \in \mathcal{T}_1} \frac{1}{M} \sum_{m=1}^M \mathbb{E} \left( \boldsymbol{\gamma}_{m^*}^{\prime} \mathbf{e}_{\boldsymbol{\mu}}^{(i,t)}  \mathbf{e}_{\boldsymbol{\mu}}^{(j,t) \prime} \boldsymbol{\gamma}_{m^*} \right) \right.\nonumber\\
&&\left. - \frac{1}{T_1} \sum_{t_1 \in \mathcal{T}_1} \frac{1}{M} \sum_{m=1}^M \mathbb{E} \left( \boldsymbol{\gamma}_{m}^{\prime} \mathbf{e}_{\boldsymbol{\mu}}^{(i,t)} \mathbf{e}_{\boldsymbol{\mu}}^{(j,t) \prime} \boldsymbol{\gamma}_{m} \right) \right| \nonumber\\
& = & \sum_{i=2}^{J+1} \sum_{j=2}^{J+1} \left| \mathbb{E} \left\{ \operatorname{tr} \left( \frac{1}{T_1} \sum_{t_1 \in \mathcal{T}_1} \frac{1}{M} \sum_{m=1}^M  \mathbf{e}_{\boldsymbol{\mu}}^{(i,t)}  \mathbf{e}_{\boldsymbol{\mu}}^{(j,t) \prime} \boldsymbol{\gamma}_{m^*} \boldsymbol{\gamma}_{m^*}^{\prime} \right) \right.\right. \nonumber\\
&& \left.\left.- \operatorname{tr} \left( \frac{1}{T_1} \sum_{t_1 \in \mathcal{T}_1} \frac{1}{M} \sum_{m=1}^M \mathbf{e}_{\boldsymbol{\mu}}^{(i,t)}  \mathbf{e}_{\boldsymbol{\mu}}^{(j,t) \prime} \boldsymbol{\gamma}_{m} \boldsymbol{\gamma}_{m}^{\prime} \right) \right\} \right| \nonumber\\
& = & \sum_{i=2}^{J+1} \sum_{j=2}^{J+1} \left| \mathbb{E} \left\{ \operatorname{tr} \left(\frac{1}{T_1} \sum_{t_1 \in \mathcal{T}_1} \Big(\frac{1}{M} \sum_{m=1}^M  (\boldsymbol{\gamma}_{m^*} \boldsymbol{\gamma}_{m^*}^{\prime} - \boldsymbol{\gamma}_{m} \boldsymbol{\gamma}_{m}^{\prime} ) \Big) \mathbf{e}_{\boldsymbol{\mu}}^{(i,t)}  \mathbf{e}_{\boldsymbol{\mu}}^{(j,t) \prime} \right) \right\} \right| \nonumber\\
& = & O(n^{-1/2} J^2), \label{e.4}
\end{eqnarray}

Finally, We consider $\mathbb{E}_3$. Same as $\mathbb{I}_3$, we have $\mathbb{E}_3=0$.

Next, we introduce relaxed versions of Assumption 2.3$^{\prime\prime}$ and Assumption 2.4$^{\prime\prime}$, together with an additional regularity
condition, and show that the asymptotic optimality result remains valid under these assumptions.

\noindent\textbf{Assumption E.1.} 
{\it
\it $\mathbb{E} (Y_{i,t-1,N} Y_{j,t-1,N} - T_1^{-1} \sum_{t_1 \in \mathcal{T}_{1}} Y_{i,t_1-1,N} Y_{j,t_1-1,N}) = o(1)$ for $t \in \mathcal{T}_0$ and $i,j \in \{1,\ldots, J+1\}$.
}

\noindent\textbf{Assumption E.2.} 
{\it
$\mathbb{E} (X_{it,l}X_{jt,l} - T_1^{-1} \sum_{t_1 \in \mathcal{T}_{1}} X_{it_1,l}X_{jt_1,l}) = o(1)$ for $t \in \mathcal{T}_0$, $l\in\{1, \ldots, p\}$ and $i,j \in \{1,\ldots, J+1\}$.
}

\noindent\textbf{Assumption E.3.} 
{\it
Let \(\epsilon_n\) be any sequence satisfying \(\epsilon_n=o(1)\). Assume that $\xi_t^{-1} J^2 \epsilon_n = o(1)$ for \(t\in\mathcal T_0\).
}

Assumption E.1 weakens Assumption 2.3$^{\prime\prime}$ by allowing the difference between the pre-treatment and post-treatment average second moments of the lagged untreated outcomes to be asymptotically negligible, rather than exactly zero. Similarly, Assumption E.2 relaxes Assumption 2.4$^{\prime\prime}$ by requiring only asymptotic stability of the second-moment structure of the covariates. Assumption E.3 is identical to Assumption D.2 and plays a role analogous to that of Assumption 1 in the proof of asymptotic optimality.

Under Assumption 1, Assumptions E.1-E.3 and (\ref{e.2})-(\ref{e.4}), we have that, for each $t \in \mathcal{T}_0$, 
\begin{eqnarray*}
&& \sup _{\mathbf{w}_t\in \mathcal{H}} \left| \frac{R_{t}(\mathbf{w}_{t}) - \bar{R}_{T_1}(\mathbf{w}_{t})} {R_{t}(\mathbf{w}_{t})} \right| \\
&\leq & \xi_{t}^{-1} \sup _{\mathbf{w}_{t} \in \mathcal{H}} \left| R_{t}(\mathbf{w}_{t}) - \bar{R}_{T_1}(\mathbf{w}_{t}) \right| \\
& \leq & \xi_{t}^{-1} \left(\mathbb{E}_1 + \mathbb{E}_2 + \mathbb{E}_3 \right) \\
&\leq & \xi_{t}^{-1} \sum_{i=2}^{J+1} \sum_{j=2}^{J+1} \left| \mathbb{E} \left\{ \operatorname{tr} \left( \frac{1}{M} \sum_{m=1}^M (\mathbf{e}_{\boldsymbol{\mu}}^{(i,t)} \mathbf{e}_{\boldsymbol{\mu}}^{(j,t) \prime} - \mathbf{e}_{\boldsymbol{\mu}}^{(i,t_1)} \mathbf{e}_{\boldsymbol{\mu}}^{(j,t_1) \prime}) \boldsymbol{\gamma}_{m^*} \boldsymbol{\gamma}_{m^*}^{\prime} \right) \right\} \right| + O\left(\xi_{t}^{-1} n^{-1/2} J^2 \right)\\
&=& o(1).
\end{eqnarray*}
This equation leads to (\ref{B3}), which, together with the analysis of (\ref{B2}), completes the proof of asymptotic optimality.

\section{Extension using mixtures of distribution functions}
In this section, we present a natural extension of the DSC method in which the mixtures of quantile functions are replaced by mixtures of distribution functions, following the approach proposed by \citet{van2024return}. We also state the asymptotic optimality of this extended method, with detailed results.

This alternative is particularly appealing when the outcome variable is discrete. 
Unlike quantile mixtures, which may produce awkward results in such settings, mixing distribution functions preserves the inherent ordinal structure of the data, making the method more suitable for categorical or ordinal outcomes.
The 2-Wasserstein distance, based on quantile mixtures, is well suited for continuous variables, as it captures smooth numerical differences across distributions by comparing their quantile functions. In contrast, the 1-Wasserstein, distance based on mixtures of distribution functions, is more appropriate for ordinal variables, which have inherent order but lack meaningful numerical intervals. For instance, in the analysis of discrete ordinal variables such as employee seniority levels (\citeauthor{van2024return}, \citeyear{van2024return}), the 1-Wasserstein distance more accurately reflects changes in the distribution across ordered categories. It is less affected by the non-existent ``numerical gaps" between ordinal levels and can accurately represent the shifts in the proportion across ordered levels.

We now describe the implementation of this synthetic control procedure based on mixtures of distribution functions.
Let $\widehat F_{Y_{jt}}(y)$ denote the empirical cumulative distribution function (CDF) of unit $j$ at time $t$, constructed from the observed outcomes $\{Y_{l, j t}\}_{l=1}^{n}$, assuming  equal sample sizes $n$ across units for simplicity. For each $t\in\mathcal{T}_0$, we determine the weights by solving 
$$\widehat{\mathbf{w}}_{t}^d  = \underset{\mathbf{w}_{t} \in \mathcal{H}}{\operatorname{argmin}} L_t^d(\mathbf{w}_t) = \underset{\mathbf{w}_{t} \in \mathcal{H}}{\operatorname{argmin}} \int_{\mathbb R}\Big|\sum_{j=2}^{J+1} w_{jt} \widehat F_{Y_{jt}}(y) - \widehat F_{Y_{1t}}(y)\Big| dy,$$
where $L_t^d(\mathbf{w}_t) = \int_{\mathbb R}\Big|\sum_{j=2}^{J+1} w_{jt} \widehat F_{Y_{jt}}(y) - \widehat F_{Y_{1t}}(y)\Big| dy$
is the 1-Wasserstein loss at time $t$, and the corresponding risk function is defined as  $R^d_t(\mathbf{w}_t) = \mathbb{E}[L^d_t(\mathbf{w}_t)]$.
We then calculate the weight $\widehat{\mathbf{w}}^d$ by a weighted average of the weights $\widehat{\mathbf{w}}_{t}^d$ over all pre-treatment periods, that is,
$$\widehat{\mathbf{w}}^d = \sum_{t \in \mathcal{T}_0} \lambda_{t} \widehat{\mathbf{w}}_{t}^d \quad \text{for } \lambda_{t}\geq 0 \text{ and } \sum_{t\in \mathcal{T}_0} \lambda_{t} = 1.$$
Finally, we construct the counterfactual CDF for the treated unit in each $t \in \mathcal{T}_1$ by 
$$\widehat{F}_{Y_{1t, N}}(y) = \sum_{j=2}^{J+1} \widehat{w}_{j}^d \widehat{F}_{Y_{j t}}(y).$$

In summary, the algorithm for this case is shown in Algorithm 2.

\begin{algorithm}
	\renewcommand{\algorithmicrequire}{\textbf{Input:}}
	\caption{DSC based on mixtures of distribution functions.}
	\label{alg:1}
	\begin{algorithmic}[1]
		\REQUIRE 1. data $Y_{l,j t}$ with $l=1, \ldots, n, j=1, \ldots, J+1, t=1, \ldots, T$\\
                 ~~~~~~2. weights $\left\{\lambda_{t}\right\}_{t \in \mathcal{T}_0}$ that satisfy $\lambda_{t}\geq 0$ and $\sum_{t \in \mathcal{T}_0} \lambda_{t} = 1$
		\STATE \textbf{procedure}
		\FOR{each time period $t \in \mathcal{T}_{0} \cup \mathcal{T}_{1}$}
        \FOR{each unit $j=1, \ldots, J+1$}
		\STATE estimate the empirical CDF $\widehat{F}_{Y_{j t}}(y)$
        \ENDFOR
		\ENDFOR
        \FOR{each time period $t \in \mathcal{T}_0$}
		\STATE obtain the weights $\widehat{\mathbf{w}}_{t}^d$ via  
        $$\widehat{\mathbf{w}}_{t}^d = \underset{\mathbf{w}_{t} \in \mathcal{H}}{\operatorname{argmin}} \int_{\mathbb R}\Big|\sum_{j=2}^{J+1} w_{jt} \widehat F_{Y_{jt}}(y) - \widehat F_{Y_{1t}}(y)\Big| dy$$
        \ENDFOR
        \STATE obtain the weight $\widehat{\mathbf{w}}^d = \sum_{t \in \mathcal{T}_0} \lambda_{t} \widehat{\mathbf{w}}_{t}^d$ over all $t \in \mathcal{T}_0$
        \FOR{each time period $t\in \mathcal{T}_1$}
		\STATE obtain the estimation of the counterfactual distribution function $\widehat{F}_{Y_{1t, N}}(y) = \sum_{j=2}^{J+1} \widehat{w}_{j}^d \widehat{F}_{Y_{j t}}(y)$
        \ENDFOR
		\STATE \textbf{end procedure}
	\end{algorithmic}
\end{algorithm}


Denote $e_{t,l}^{(j)} = \mathbb{I}\{Y_{l,jt} \leq y \} - \mathbb{I}\{Y_{l,1t} \leq y \}$ for $j = 2, \ldots, J+1$, $l = 1, 2, \ldots, n$ and $t \in \mathcal{T}_0 \cup \mathcal{T}_1$. To establish the asymptotic optimality, we impose the following assumptions.

\noindent\textbf{Assumption F.1.}~~{\it There exists a constant $C_5>0$ such that, for any $j \in\{2, \ldots, J+1\}$ and $t \in \mathcal{T}_0 \cup \mathcal{T}_1$, $\operatorname{var} (n^{-1 / 2} \sum_{l=1}^n e_{t, l}^{(j)}) \geq C_5$.
}

\noindent\textbf{Assumption F.2.}~~{\it $\sup_{y \in \mathbb R} \frac{1}{T_1} \sum_{s \in \mathcal{T}_1} | F_{Y_{jt}}(y) - F_{Y_{js}} (y)| = o(1)$ for any $j=1, \ldots, J+1$ and $t \in \mathcal{T}_0$.
}

\noindent\textbf{Assumption F.3.}~~{\it
Let \(\epsilon_n\) be any sequence satisfying \(\epsilon_n=o(1)\). Assume that $\xi_t^{-1} J \epsilon_n = o(1)$ for \(t\in\mathcal T_0\).
}

Assumption F.1 is a non-degeneracy condition that rules out vanishing variability in the indicator differences, ensuring that the variance of these differences remains bounded away from zero as $n$ increases. 
Assumption F.2 is a uniform stability condition requires that, for each unit, the post-treatment distributions remain uniformly close to their pre-treatment counterpart. It guarantees that the weight vector chosen based on pre-treatment data can approximate post-treatment distributions with comparable accuracy. 
Assumption F.3 regulates the joint growth of the number of units \(J\) and \(\xi_t^{-1}\) relative to the convergence rate \(\epsilon_n\). Intuitively, even if \(J\) is large or \(\xi_t\) is small, their product must not increase too fast relative to any vanishing sequence \(\epsilon_n\). This assumption is analogous to Assumption 1, and similarly implies that $\xi_{t} \neq 0$.

To evaluate post-treatment fit, we consider the average 1-Wasserstein distance, defined as
$$\bar{R}^d_{T_1}(\mathbf{w}) = \frac{1}{T_1} \sum_{t \in \mathcal{T}_1} \int_{\mathbb R}\Big|\sum_{j=2}^{J+1} w_j F_{Y_{jt}}(y)-F_{Y_{1t, N}}(y)\Big| dy,$$
where $F_{Y_{jt}}$ denotes the true outcome distribution for unit $j$ at time $t$. 

\begin{theorem}\label{th3}
\it Given any $\mathbf{\lambda}$, if $T_0$ is finite, then under Assumptions F.1-F.3, we have
\begin{align}
\frac{\bar{R}^d_{T_1}(\widehat{\mathbf{w}}^d)}{\inf _{\mathbf{w} \in \mathcal{H}} \bar{R}^d_{T_1}(\mathbf{w})} \stackrel{p}{\rightarrow} 1. \label{E1}
\end{align}
\end{theorem}

\noindent{\it Proof of Theorem 4.} We follow the proof of Theorem 1 to verify Theorem 4. Based on the analysis in Appendix A, to prove (\ref{E1}), it suffices to show that
\begin{align}
\sup _{\mathbf{w}_{t^{\prime}} \in \mathcal{H}} \left| \frac{L^d_{t^{\prime}}(\mathbf{w}_{t^{\prime}}) - R^d_{t^{\prime}} (\mathbf{w}_{t^{\prime}})} {R^d_{t^{\prime}}(\mathbf{w}_{t^{\prime}})} \right| = o_p(1) \label{E2}
\end{align}
and
\begin{align}
\sup _{\mathbf{w}_{t^{\prime}} \in \mathcal{H}} \left| \frac{R^d_{t^{\prime}}(\mathbf{w}_{t^{\prime}}) - \bar{R}^d_{T_1}(\mathbf{w}_{t^{\prime}})} {R^d_{t^{\prime}}(\mathbf{w}_{t^{\prime}})} \right| = o(1). \label{E3}
\end{align}

First, we give the proof of (\ref{E2}). Note that
\begin{eqnarray}
&&\sup _{\mathbf{w}_{t^{\prime}} \in \mathcal{H}} \left| \frac{L^d_{t^{\prime}}(\mathbf{w}_{t^{\prime}}) - R^d_{t^{\prime}} (\mathbf{w}_{t^{\prime}})} {R^d_{t^{\prime}}(\mathbf{w}_{t^{\prime}})} \right| \notag\\
&\leq & \xi_{t^{\prime}}^{-1} \sup _{\mathbf{w}_{t^{\prime}} \in \mathcal{H}} \left| L^d_{t^{\prime}}(\mathbf{w}_{t^{\prime}}) - R^d_{t^{\prime}} (\mathbf{w}_{t^{\prime}}) \right| \notag\\
&=& \xi_{t^{\prime}}^{-1} \sup _{\mathbf{w}_{t^{\prime}} \in \mathcal{H}} \left| \int_{\mathbb R}\Big|\sum_{j=2}^{J+1} w_{j t^{\prime}} \widehat F_{Y_{jt^{\prime}}}(y) - \widehat F_{Y_{1t^{\prime}}}(y)\Big| dy - \mathbb{E} \left[\int_{\mathbb R} \Big|\sum_{j=2}^{J+1} w_{j t^{\prime}} \widehat F_{Y_{jt^{\prime}}}(y) - \widehat F_{Y_{1t^{\prime}}}(y)\Big| dy \right] \right| \notag\\
&\leq& \xi_{t^{\prime}}^{-1} \sup _{\mathbf{w}_{t^{\prime}} \in \mathcal{H}} \sum_{j=2}^{J+1} | w_{j t^{\prime}} | \left| \int_{\mathbb R} \left( \Big| \widehat F_{Y_{jt^{\prime}}}(y) - \widehat F_{Y_{1t^{\prime}}}(y)\Big| - \mathbb{E} \Big|\widehat F_{Y_{jt^{\prime}}}(y) - \widehat F_{Y_{1t^{\prime}}}(y)\Big| \right) dy \right| \notag\\
&\leq& \xi_{t^{\prime}}^{-1} \sum_{j=2}^{J+1} \left| \int_{\mathbb R} \left( \Big| \widehat F_{Y_{jt^{\prime}}}(y) - \widehat F_{Y_{1t^{\prime}}}(y)\Big| - \mathbb{E} \Big| \widehat F_{Y_{jt^{\prime}}}(y) - \widehat F_{Y_{1t^{\prime}}}(y)\Big| \right) dy \right| \notag\\
&=& \xi_{t^{\prime}}^{-1} \sum_{j=2}^{J+1} \left| \int_{\mathbb R} \left( \Big| \frac{1}{n} \sum_{l=1}^{n} \mathbb{I}\{Y_{l,jt^{\prime}} \leq y \} - \frac{1}{n} \sum_{l=1}^{n} \mathbb{I}\{Y_{l,1t^{\prime}} \leq y \} \Big| \right.\right. \notag\\
&& \left.\left. - \mathbb{E} \Big| \frac{1}{n} \sum_{l=1}^{n} \mathbb{I}\{Y_{l,jt^{\prime}} \leq y \} - \frac{1}{n} \sum_{l=1}^{n} \mathbb{I}\{Y_{l,1t^{\prime}} \leq y \} \Big| \right) dy \right| \notag\\
&\leq& \xi_{t^{\prime}}^{-1} \sum_{j=2}^{J+1} \left|\frac{1}{n} \sum_{l=1}^n \int_{\mathbb R} \left( \Big| \mathbb{I}\{Y_{l,jt^{\prime}} \leq y \} - \mathbb{I}\{Y_{l,1t^{\prime}} \leq y \} \Big| - \mathbb{E} \Big| \mathbb{I}\{Y_{l,jt^{\prime}} \leq y \} - \mathbb{I}\{Y_{l,1t^{\prime}} \leq y \} \Big| \right) dy \right| \notag\\
&=& \xi_{t^{\prime}}^{-1} \sum_{j=2}^{J+1} {n}^{-1 / 2} \Psi_{l, t^{\prime}}(j), \label{E4}
\end{eqnarray}
where
\begin{align*}
\Psi_{l, t^{\prime}}(j) = \left|\frac{1}{\sqrt{n}} \sum_{l=1}^n \int_{\mathbb R} \left( \Big| \mathbb{I}\{Y_{l,jt^{\prime}} \leq y \} - \mathbb{I}\{Y_{l,1t^{\prime}} \leq y \} \Big| - \mathbb{E} \Big| \mathbb{I}\{Y_{l,jt^{\prime}} \leq y \} - \mathbb{I}\{Y_{l,1t^{\prime}} \leq y \} \Big| \right) dy \right|.
\end{align*}
Based on Assumption F.1, we can obtain that
\begin{align}
\sum_{j=2}^{J+1} \Psi_{l, t^{\prime}}(j) = O_p (J). \label{E5}
\end{align}
Combining (\ref{E4}), (\ref{E5}) and Assumption 1, we can obtain (\ref{E2}).

Then, we show that (\ref{E3}) holds.
\begin{eqnarray}
&& \sup_{\mathbf{w}_{t^{\prime}} \in \mathcal{H}} \left| \frac{R^d_{t^{\prime}}(\mathbf{w}_{t^{\prime}}) - \bar{R}^d_{T_1}(\mathbf{w}_{t^{\prime}})} {R^d_{t^{\prime}}(\mathbf{w}_{t^{\prime}})} \right| \notag\\
&\leq & \xi_{t^{\prime}}^{-1} \sup _{\mathbf{w}_{t^{\prime}} \in \mathcal{H}} \left| R^d_{t^{\prime}}(\mathbf{w}_{t^{\prime}}) - \bar{R}^d_{T_1}(\mathbf{w}_{t^{\prime}}) \right| \notag\\
&=& \xi_{t^{\prime}}^{-1} \sup _{\mathbf{w}_{t^{\prime}} \in \mathcal{H}} \left| \mathbb{E} \left[\int_{\mathbb R} \Big|\sum_{j=2}^{J+1} w_{j t^{\prime}} \widehat F_{Y_{jt^{\prime}}}(y) - \widehat F_{Y_{1t^{\prime}}}(y)\Big| dy \right] \right. \notag\\
&&\left. - \frac{1}{T_1} \sum_{t\in \mathcal{T}_1} \int_{\mathbb R} \Big|\sum_{j=2}^{J+1} w_{j t^{\prime}} F_{Y_{jt}}(y) - F_{Y_{1t, N}}(y)\Big| dy \right| \notag\\
&\leq& \xi_{t^{\prime}}^{-1} \sup _{\mathbf{w}_{t^{\prime}} \in \mathcal{H}} \left\{\left| \mathbb{E} \left[\int_{\mathbb R} \Big|\sum_{j=2}^{J+1} w_{j t^{\prime}} \widehat F_{Y_{jt^{\prime}}}(y) - \widehat F_{Y_{1t^{\prime}}}(y)\Big| dy \right] - \int_{\mathbb R} \Big|\sum_{j=2}^{J+1} w_{j t^{\prime}} F_{Y_{jt^{\prime}}}(y) - F_{Y_{1t^{\prime}}}(y)\Big| dy \right|\right. \notag\\
&&+ \left.\left| \int_{\mathbb R} \Big|\sum_{j=2}^{J+1} w_{j t^{\prime}} F_{Y_{jt^{\prime}}}(y) - F_{Y_{1t^{\prime}}}(y)\Big| dy - \frac{1}{T_1} \sum_{t\in \mathcal{T}_1} \int_{\mathbb R} \Big|\sum_{j=2}^{J+1} w_{j t^{\prime}} F_{Y_{jt}}(y) - F_{Y_{1t, N}}(y)\Big| dy \right|\right\} \notag\\
&\leq& \xi_{t^{\prime}}^{-1} \sum_{j=2}^{J+1} \left| \int_{\mathbb R} \mathbb{E} \Big|\widehat F_{Y_{jt^{\prime}}}(y) - \widehat F_{Y_{1t^{\prime}}}(y)\Big| dy - \int_{\mathbb R} \Big| F_{Y_{jt^{\prime}}}(y) - F_{Y_{1t^{\prime}}}(y)\Big| dy \right| \notag\\
&& + \xi_{t^{\prime}}^{-1} \sum_{j=2}^{J+1} \left| \int_{\mathbb R} \Big| F_{Y_{jt^{\prime}}}(y) - F_{Y_{1t^{\prime}}}(y)\Big| dy - \frac{1}{T_1} \sum_{t\in \mathcal{T}_1} \int_{\mathbb R} \Big| F_{Y_{jt}}(y) - F_{Y_{1t, N}}(y)\Big| dy \right| \notag\\
& \equiv &\mathbb{A}_1 + \mathbb{A}_2. \label{E6}
\end{eqnarray}

We begin by analyzing $\mathbb{A}_1$. By Donsker’s theorem and the Glivenko–Cantelli property, it holds that
$$
\sup_{y\in{\mathbb R}} \mathbb{E} |\widehat F_{Y_{jt}}(y) - F_{Y_{jt}}(y)| = O(n^{-1/2}),
$$
uniformly in $j$ and $t$. Combined with some elementary inequalities, we can then derive that
\begin{eqnarray}
\mathbb{A}_1 & = & \xi_{t^{\prime}}^{-1} \sum_{j=2}^{J+1} \left| \int_{\mathbb R} \left(\mathbb{E} \Big|\widehat F_{Y_{jt^{\prime}}}(y) - \widehat F_{Y_{1t^{\prime}}}(y)\Big| - \Big| F_{Y_{jt^{\prime}}}(y) - F_{Y_{1t^{\prime}}}(y)\Big| \right) dy \right| \nonumber\\
& \leq & \xi_{t^{\prime}}^{-1} \sum_{j=2}^{J+1} \int_{\mathbb R} \mathbb{E} \left| |\widehat F_{Y_{jt^{\prime}}}(y) - \widehat F_{Y_{1t^{\prime}}}(y) | - | F_{Y_{jt^{\prime}}}(y) - F_{Y_{1t^{\prime}}}(y) | \right| dy  \nonumber\\
& \leq & \xi_{t^{\prime}}^{-1} \sum_{j=2}^{J+1} \int_{\mathbb R} \mathbb{E} \left| (\widehat F_{Y_{jt^{\prime}}}(y) - \widehat F_{Y_{1t^{\prime}}}(y) ) - (F_{Y_{jt^{\prime}}}(y) - F_{Y_{1t^{\prime}}}(y)) \right| dy  \nonumber\\
& \leq & \xi_{t^{\prime}}^{-1} \sum_{j=2}^{J+1} \int_{\mathbb R} \left(\mathbb{E} | \widehat F_{Y_{jt^{\prime}}}(y) - F_{Y_{jt^{\prime}}}(y) | + \mathbb{E} |\widehat F_{Y_{1t^{\prime}}}(y) - F_{Y_{1t^{\prime}}}(y)| \right) dy  \nonumber\\
& = & O(\xi_{t^{\prime}}^{-1} n^{-1/2} J). \label{E7}
\end{eqnarray}
Together with Assumption 1, we conclude that $\mathbb{A}_1 = o(1)$. We now turn to the analysis of $\mathbb{A}_2$. From Assumptions F.2 and F.3, we have that
\begin{eqnarray}
\mathbb{A}_2 & = & \xi_{t^{\prime}}^{-1} \sum_{j=2}^{J+1} \left| \int_{\mathbb R} \Big| F_{Y_{jt^{\prime}}}(y) - F_{Y_{1t^{\prime}}}(y)\Big| dy - \frac{1}{T_1} \sum_{t\in \mathcal{T}_1} \int_{\mathbb R} \Big| F_{Y_{jt}}(y) - F_{Y_{1t, N}}(y)\Big| dy \right| \nonumber\\
& \leq & \xi_{t^{\prime}}^{-1} \sum_{j=2}^{J+1} \frac{1}{T_1} \sum_{t\in \mathcal{T}_1} \left| \int_{\mathbb R} \left(| F_{Y_{jt^{\prime}}}(y) - F_{Y_{1t^{\prime}}}(y) | - | F_{Y_{jt}}(y) - F_{Y_{1t, N}}(y) | \right)dy \right| \nonumber\\
& \leq & \xi_{t^{\prime}}^{-1} \sum_{j=2}^{J+1} \frac{1}{T_1} \sum_{t\in \mathcal{T}_1} \int_{\mathbb R} \left| ( F_{Y_{j t^{\prime}}}(y) - F_{Y_{1 t^{\prime}}}(y)) - (F_{Y_{jt}}(y) - F_{Y_{1t, N}}(y)) \right|dy \nonumber\\
& \leq & \xi_{t^{\prime}}^{-1} \sum_{j=2}^{J+1} \frac{1}{T_1} \sum_{t\in \mathcal{T}_1} \int_{\mathbb R} \left( | F_{Y_{j t^{\prime}}}(y) - F_{Y_{jt}}(y) | + |F_{Y_{1 t^{\prime}}}(y) - F_{Y_{1t, N}}(y) | \right) dy \nonumber\\
& = & o(1). \label{E8}
\end{eqnarray}
This completes the proof of Theorem 4.

\bibliographystyle{apalike}
\bibliography{refs}

\end{document}